\documentclass[english]{article}
\usepackage[latin9]{inputenc}
\usepackage{geometry}
\usepackage{color}
\usepackage{babel}
\usepackage{array}
\usepackage{url}
\usepackage{multirow}
\usepackage{amsmath}
\usepackage{amssymb}
\usepackage{graphicx}
\usepackage[unicode=true,pdfusetitle,
 bookmarks=true,bookmarksnumbered=false,bookmarksopen=false,
 breaklinks=false,pdfborder={0 0 1},backref=false,colorlinks=false]
 {hyperref}

\usepackage{epstopdf}

\makeatletter

\providecommand{\tabularnewline}{\\}

\usepackage{wrapfig,psfig,epsf,times}

\makeatother

\begin{document}

\title{Stabilized Sparse Ordinal Regression for Medical Risk Stratification}

\author{Truyen Tran,  Dinh Phung,  Wei Luo, Svetha Venkatesh\\
Center for Pattern Recognition and Data Analytics (PRaDA)\\
Deakin University, Australia.\\
\{truyen.tran, dinh.phung,wei.luo, svetha.venkatesh\}@deakin.edu.au
}

\date{}

\maketitle
\begin{abstract}
The recent wide adoption of Electronic Medical Records (EMR) presents
great opportunities and challenges for data mining. The EMR data is
largely temporal, often noisy, irregular and high dimensional. This
paper constructs a novel ordinal regression framework for predicting
medical risk stratification from EMR. First, a conceptual view of
EMR as a temporal image is constructed to extract a diverse set of
features. Second, ordinal modeling is applied for predicting cumulative
or progressive risk. The challenges are building a transparent predictive
model that works with a large number of weakly predictive features,
and at the same time, is stable against resampling variations. Our
solution employs sparsity methods that are stabilized through domain-specific
feature interaction networks. We introduces two indices that measure
the model stability against data resampling. Feature networks are
used to generate two multivariate Gaussian priors with sparse precision
matrices (the Laplacian and Random Walk). We apply the framework on
a large short-term suicide risk prediction problem and demonstrate
that our methods outperform clinicians to a large-margin, discover
suicide risk factors that conform with mental health knowledge, and
produce models with enhanced stability.
\end{abstract}
\global\long\def\LL{\mathcal{L}}
\global\long\def\Data{\mathcal{D}}
\global\long\def\Wb{\boldsymbol{W}}

\global\long\def\zb{\boldsymbol{z}}
\global\long\def\wb{\boldsymbol{w}}
\global\long\def\xb{\boldsymbol{x}}
\global\long\def\fb{\boldsymbol{f}}

\global\long\def\bx{\boldsymbol{x}}
\global\long\def\bX{\boldsymbol{X}}
\global\long\def\bW{\mathbf{W}}
\global\long\def\bH{\mathbf{H}}
\global\long\def\bL{\mathbf{L}}
\global\long\def\tbx{\tilde{\bx}}
\global\long\def\by{\boldsymbol{y}}

\global\long\def\bY{\boldsymbol{Y}}
\global\long\def\bz{\boldsymbol{z}}
\global\long\def\bZ{\boldsymbol{Z}}
\global\long\def\bu{\boldsymbol{u}}
\global\long\def\bU{\boldsymbol{U}}
\global\long\def\bv{\boldsymbol{v}}
\global\long\def\bV{\boldsymbol{V}}

\section{Introduction\label{sec:Introduction}}

The recent wide adoption of Electronic Medical Records (EMRs) offers
great opportunities for mining useful patterns that support clinical
research and decision making \cite{jensen2012mining}. The EMR contains
rich information about a patient, including demographics, history
of hospital visits, diagnoses, physiological measurements, bio-markers
and interventions. We consider the problem of predicting risk stratification
using EMR data. By `risk' we mean unwanted outcomes such as readmissions,
length of hospitalization, intoxication and mortality. For clinical
use, the outcomes are often stratified into ordered levels such as
``low'', ``moderate'' and ``high'' risk. We aim at constructing
a scalable automated framework that takes entire historical medical
records for each patient and predicts ordered risk within a window. 

The challenges lie in effective and interpretable modeling of noisy,
irregular, temporal and mixed modalities \cite{luo2012sor,wang2012towards}.
An EMR can be considered as a mixture of static information and time-stamped
events. Static information includes demographic variables and thus
is generally moderate in dimensions. The events are, however, complex
and high dimensional. For example, the current disease coding scheme
ICD-10 has approximately $20,000$ entries, and the number adds up
quickly if we consider multiple time scales and the combination with
other event types (e.g., medications). Events are often packed into
episodes of admissions and treatments, and thus are highly irregular.

The high dimensional data calls for sparse predictive models \cite{tibshirani1996rsa,ye2012sparse}.
Unfortunately, sparse models could be unstable against data variations.
The instability can be measured as the probability that a feature
is selected \cite{meinshausen2010stability} or as the variance in
model parameters. In EMR, features can be highly correlated, and thus
sparse models often pick the strongest one seen in the data sample
\cite{xu2011sparse,zou2005regularization}. Under data resampling,
another feature may be chosen next time. Second, for some tasks, EMR-derived
features could be weakly predictive, thus limiting the probability
that they are selected. Unstable models are less useful in practice
because they cannot generalize from one cohort to another, and thus
undermining the research reproducibility and reducing the clinician
adoption.

In this paper, we present a two-stage \emph{stabilized sparse} \emph{ordinal}
framework that addresses these challenges \cite{tran2013integrated}.
The first stage extracts a large number of features from EMR. The
extraction builds on a novel conceptual view in which a patient's
medical records forms a temporal image, from which filters over different
time windows extract a diverse feature set. Multi-class ordinal classification
is then formulated in two main ways---with and without class-specific
parameters. For each set, risk is transparently modeled as either
cumulative risk \cite{mccullagh1980rmo} or stagewise progression
of risk \cite{tutz1991sequential,Truyen:2012b}. The ordinal classifiers
are equipped a $\ell_{1}$-norm penalty which yields sparse solutions.
The selection is stabilized through \emph{relational regularization},
in which domain-specific feature interaction is used to promote smoothness
among related parameters. Examples of interaction include the ``sibling''
relation between two diagnosis codes in the same disease branch, or
the progression of a disease. To measure model stability, we introduce
two stability indices, evaluated at any feature ranked list length,
one accounts for the feature \emph{selection probability} and the
other computes the \emph{signal-to-noise ratio} in the feature weights.

Our framework is demonstrated through a large cohort of ten of thousands
of mental health patients who were under assessment for suicide risk.
This problem devastates families and communities: One out of ten persons
develop suicidal thoughts in their lifetime \cite{nock2013prevalence},
and 0.3 percents attempt suicide in any given year \cite{borges2010twelve}.
In response, health services introduced mandatory suicide risk assessment
for vulnerable populations \cite{allen2013screening}. These assessments
form the basis of suicide risk stratification. But traditional suicide
risk assessment lacks prediction accuracy \cite{large2011validity,ryan2010clinical}.
Providing more accurate solutions for suicide risk stratification
will deliver immediate benefits. The challenges are that data is highly
sparse; many risk factors are known, but they are weakly predictive. 

The framework is evaluated against several criteria: predictive accuracy
against clinicians, the degree to which discovered features conform
with clinical knowledge, and model stability. In predicting risk,
the framework outperforms the mental health professionals in a large
margin. For moderate-risk prediction, machines improve the $F_{1}$-score
by $25\%$. For the high-risk class, the improvement are as high as
$200\%$. In terms of suicide detection, the machine detects $29-30$
cases, which are more than double the number detected by human ($14$
cases). The discovered features agree with most previously reported
risk factors which came out of decades of extensive research. The
results are significant as the framework relies entirely on data readily
collected in hospitals, and the risk prediction is objective and transparent.
We also demonstrate the efficacy of our feature stabilization methods
vs no stabilization. 

In short, this paper contains the following contributions:
\begin{itemize}
\item A generic and scalable risk stratification framework with three components:
(i) A novel conceptual view of EMR as a temporal image so that a diverse
set of features at different temporal scales can be extracted; (ii)
Modeling of risk through ordinal classification, in particular the
introduction of stagewise risk modeling in this context; (iii) Formulation
of methods to stabilize predictive models, through appropriate relational
regularization of the risk functional in ordinal classification.
\item Two model stability indices over arbitrary feature rank list size,
one is based on probability that a feature is selected, and the other
on signal-to-noise ratio of the feature weights. A related contribution
is a novel stability-based feature ranking criterion based on the
signal-to-noise ratios. 
\item Comprehensive evaluation of methods in comparison with clinicians,
demonstrating that machine learning methods outperform clinicians
in risk stratification. It demonstrates the value of mining EMR data
for an important problem. To the best our knowledge, this is the first
study that formulates suicide risk prediction as a data mining task
and leads to a solution being clinically adopted. 
\item The framework can be generalized to a variety of disease. Given mixed
type data comprising demography, clinical history (emergency attendances,
admissions and diagnostic coding), and risk assessment instruments
(questions with ordinal ratings), our framework automatically extracts
the most relevant features and builds stabilized risk prediction classifiers. 
\end{itemize}
The paper is organized as follows. The next section reviews related
work. Section~\ref{sec:Framework-Overview} presents an overview
of the framework. Section~\ref{sec:Data-Modelling} describes EMR
data representation and feature extraction. Ordinal classifiers are
derived in the subsequent section, followed by the relational stabilization
method in Section~\ref{sec:Stabilized-Prediction}. Section~\ref{sec:Implementation-and-Results}
details implementation issues and results. Section~\ref{sec:Discussion}
provides further discussion, followed by the conclusion.

\section{Background \label{sec:Background}}

Risk stratification is important in medical practice and research
\cite{steyerberg2009clinical}. There are two major prediction types:
diagnosis (estimating the probability that a disease is present) and
prognosis (predicting the outcomes given current diagnoses or intervention
plan). These estimation and prediction influence clinical practices
such as test ordering, treatment/discharge planning and resource allocation.
In medical research, knowing the risk helps selecting cohorts for
randomized trials and assessing risk aspects and confounding factors.

The established risk model construction strategy relies on small hand-picked
subset of features from highly stratified cohorts \cite{oquendo2012machine,steyerberg2009clinical}.
As a result, previous studies were fragmented where conclusion only
holds under well-controlled conditions. Electronic Medical Record
(EMR), on the other hand, suggests a data-centric and hypothesis-free
approach from which data mining techniques can be utilized. It typically
contain a diverse set of information types, including demographics,
admissions and diagnoses, lab tests and treatments. Research into
machine learning for EMRs is largely recent and fast growing \cite{jensen2012mining}.
However, automating the learning process is still limited. In our
work, we generate, select and combine thousands of weak signals in
an automated fashion.

\subsubsection*{Sparsity and Stability}

The nature of such high-dimensional setting leads to sparse models,
where only small subsets of strongly predictive signals are kept.
Such sparsity leads to better interpretability and generalization;
this is expected to play an important role in biomedical domains \cite{ye2012sparse}.
However, sparse models alone are not enough in practice. We need stable
models, that is, models that do not change significantly with data
sampling. Stable models are reproducible and generalizable from one
cohort to another. However, sparsity and stability could be conflicting
goals \cite{xu2011sparse}, especially when noise is present \cite{donoho2006stable}.
Most sparsity-inducing algorithms do not aim at producing stable models.
For example, stepwise feature selection in logistic regression produces
unstable models \cite{austin2004automated}. In the context of lasso,
only one variable is chosen if two are highly correlated \cite{zou2005regularization}. 

Model stability is related but distinct from prediction stability
-- the predictive power does not change due to small perturbation
in training data \cite{bousquet2002stability}. It is quite possible
that unstable models can still produce stable outputs. Stable models,
on the other hand, lead to stable prediction under regular conditions
often seen in practice. Model stability is a stronger requirement
than recently studied feature selection stability issues \cite{bi2003dimensionality,lausser2013measuring}.
Stable feature selection algorithms produce similar feature subsets
under data variation; whilst model stability also considers feature
weights.

Several model stability indices have been introduced recently \cite{gulgezen2009stable,kalousis2007stability,kuncheva2007stability,lausser2013measuring,somol2010evaluating}.
A popular strategy is to consider similarity between any feature set
pair, each of which could be represented using the discrete set, rank,
or a weight list. The mean similarity is then considered as the stability
of the collection of feature sets. One problem with this approach
is that the stability often increases as more features are included;
and this does not reflect the domain intuition that a small subset
of strong features should be more stable than large, weak subsets.

A common method to improve the stability is to exploit aggregated
information such as set statistics \cite{abraham2010prediction},
averaging \cite{park2007averaged} or rank aggregation (e.g., see
\cite{soneson2012framework} for a references therein). The second
approach quantifies the redundancy in the feature set, i.e., exploiting
feature exchangeability \cite{soneson2012framework}, and group-based
selection \cite{yu2008stable,yuan2006model}. 

The stabilization method introduced in this paper relies on relations
between features, i.e., similar features would have similar weights.
Since this knowledge is independent of data sampling, model variation
due to sampling noise will be reduced. Feature networks have been
previously suggested in different contexts, e.g., for regularization
or improving interpretability, but the stabilization property has
been largely ignored \cite{miguel2011network,tibshirani2005sparsity,zhou2013modeling}.
Likewise, the network-based sparsity is part of a recent body of research
known as structural sparsity \cite{huang2011learning,luo2012toward}.
For instance, when the feature network is fragmented into tightly
connected subnetworks (cliques), we yield a sparse group setting \cite{yuan2006model}.
However, our work does not primarily aim to select groups of features
but rather to improve the stability (hence reproducibility) of the
selected subset.

\subsubsection*{Ordinal Regression}

The nature of medical risk suggests the use of ordinal scales since
they naturally represent human judgment \cite{bender1997ordinal}.
The most frequently used ordinal regression model is the Proportional
Odds \cite{mccullagh1980rmo}, where the odds ratio of risk above
a level and risk below it is proportional to risk factors. This model
is a special case of the assumption that risk is cumulative, and there
is a natural grouping of continuous risk into consecutive intervals,
separated by thresholds. Similar ideas have been studied in machine
learning under the kernel methods \cite{chu2007svo,herbrich1999large,sun2010kernel,chu2006gpo,crammer2002pr}.
Kernel methods allow nonlinear modeling with well-studied generalization
bounds. However, these methods could be slow for large-scale problems
since the learning complexity could be cubic in number of training
points and the testing complexity is linear in the number of support
vectors, which could be the number of training points in the worst
case. Another approach is to reduce ordinal regression to binary classification
for which standard machine learning techniques can be applied \cite{cardoso2007learning,li2006ordinal}.

A sparse probabilistic model whose risk is linear in predictors would
scale better in testing phase, regardless of the number of training
data points. It also conforms with clinician\textquoteright{}s reasoning
strategies under uncertainty when the risk is additive and the outcomes
are stated in probability.

\subsubsection*{Medical and Suicidal Risk Stratification}

Like other medical problems, suicide risk analysis is often based
on a small number of well-chosen risk factors \cite{pokorny1983prediction}\cite{brown2000risk}.
Most clinical research, however, focuses on quantifying the risk factors
rather than building a prediction model. The most common practice
in risk assessment is using questionnaires to quantify aspects related
to suicide ideation and attempts. Although mandatory, this practice
is inadequate in predicting future suicide\emph{ }\cite{large2011validity,ryan2010clinical}\emph{.
}More recently, multiple risk assessment instruments have been combined
to improve the risk judgment \cite{blasco2012combining}.

Machine learning techniques such as SVM and neural networks applied
to clinical data is typically aimed at achieving higher predictive
performance, and thus interpretability may be sacrificed. The application
to suicide risk prediction is limited \cite{modai2002neural}. In
\cite{delgado2011improving}, authors use impulsiveness scale items
to classify attempts from non attempts. However, it is unclear that
this is prediction into the future, or just separating recorded but
ambiguous facts. The work of \cite{ruiz2012bayesian} analyzes questionnaires
to discover latent features from data. The study is limited to suicide
ideation, which is poorly related to real attempted or completed suicides
in the future. Another line of work is to analyze suicidal notes \cite{pestian2010suicide}
using NLP techniques. While this is important to understanding suicidal
drive, it may not be applicable in predicting future suicide because
notes are generally not available prior to suicidal events.

\section{Framework Overview \label{sec:Framework-Overview}}

\begin{figure*}
\begin{centering}
\includegraphics[width=0.95\textwidth]{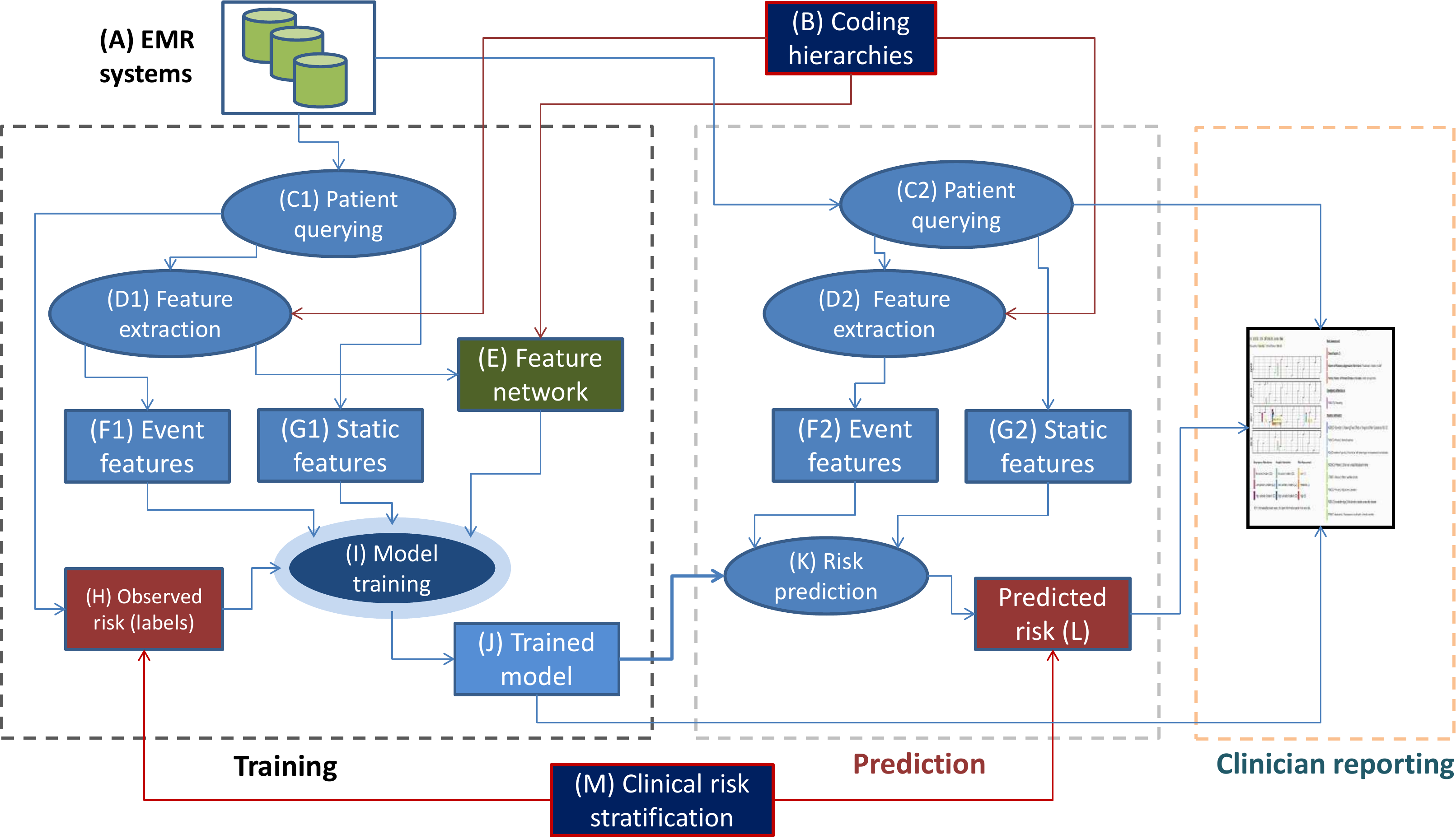}
\par\end{centering}

\centering{}\caption{Overview of the automated medical risk stratification framework. Models
are updated offline on a regular basis, prediction is made online
at clinician's request. The clinician reporting system is described
elsewhere \cite{rana2014healthmap}. \label{fig:Framework-overview}}
\end{figure*}

The framework is built on the patient-specific data queried from the
relational EMR systems (denoted as A in Fig.~\ref{fig:Framework-overview}).
Patient data contains time-stamped events (such as emergency visits,
diagnoses, and hospitalizations) and static information (such as gender,
spoken language and occupation). For each patient, there are one or
several evaluation points from which future risk will be predicted
(Fig.~\ref{fig:Event-image-filter-bank}). Often clinical risk assessments,
hospital admissions or discharges serve as natural evaluation points
as the outcomes will be tracked and acted upon. 

 The \emph{feature extraction} process (D1,D2) generates \emph{event
features} (F1,F2) over multiple periods of times prior to an evaluation
point (Section~\ref{sec:Multiscale-Feature-Extraction}). The extraction
process makes use of pre-defined \emph{coding hierarchies} (B) such
as the international disease coding scheme ICD-10%
\footnote{http://apps.who.int/classifications/icd10%
} and the Australian intervention coding scheme ACHI%
\footnote{http://www.aihw.gov.au/procedures-data-cubes%
}. In the training phase, the process also generates a \emph{feature
network} (E) which encodes the temporal and semantic relations between
features. For example, if depressive episodes were observed twice
in the history, then the two features representing them are temporally
linked. On the other hand, if another mental disorder is also observed,
then the two disorders as semantically linked. The feature network
will be used later on to improve the model stability.

The feature extraction process effectively flattens the structured
EMRs into vectors, however temporal and hierarchical information is
partially preserved. This process typically produces a large pool
of features, and thus a feature selection capacity is needed. This
is realized through model training (I) with \emph{lasso}-style regularization
\cite{tibshirani1996rsa}. More formally, let $\mathcal{D}=\left\{ \xb_{i},y_{i}\right\} _{i=1}^{n}$
be the training data set, where $\xb_{i}\in\mathbb{R}^{d}$ denotes
the feature vector of data instance $i$ and $y_{i}\in\left\{ 1,2,..,L\right\} $
the discrete ordinal output. We aim to learn a sparse, linear risk
model parameterized by the weight vector $\wb\in\mathbb{R}^{d}$.
The lasso-regularized loss function is as follows:
\begin{equation}
\mbox{loss}_{1}(\wb)=\frac{1}{n}\sum_{i=1}^{n}R(\xb_{i},y_{i};\wb)+\alpha\left\Vert \wb\right\Vert _{1}\label{eq:loss-func-lasso}
\end{equation}
where $R(\xb_{i},y_{i};\wb)$ is a convex loss function of training
instance $i$ and $\alpha>0$ is the regularization parameter. To
accommodate ordered risk classes, we employ several probabilistic
ordinal classifiers what make different assumptions about the stratification
process (Section~\ref{sec:Risk-Modelling}). The loss function $R(\xb_{i},y_{i};\wb)$
is therefore the negative log-likelihood of the outcomes $y_{i}$
given the features $\xb_{i}$.

\subsubsection*{Model Stabilization Using Feature Network}

The lasso tends to result in sparse models with few non-zeros weights.
However, we observed that this sparsity typically comes with instability
of the model under the random sampling of the training data $\mathcal{D}$.
Model instability under feature selection is indeed a known phenomenon
\cite{bi2003dimensionality}, but the theoretical study under lasso
is recent \cite{meinshausen2010stability}. 

The method proposed in this paper is based on the intuition that strong
prior knowledge would lead to less variation due to sampling noise
since prior knowledge is independent of sampling procedures. In clinical
domains, prior knowledge could be realized by using feature networks,
exploiting the relations between diseases and disease progression
over time. In Fig.~\ref{fig:Framework-overview}, the feature network
(E) links related features and ensures that similar features have
similar weights. This can be nicely formulated in a Bayesian regularization
fashion as the feature network serves as a backbone for a precision
matrix of the multivariate Gaussian prior distribution. Section~\ref{sec:Stabilized-Prediction}
presents the network regularization in more details.

\section{Representing Medical Data \label{sec:Data-Modelling}}

This section describes the data and the process of transforming the
temporal, hierarchical and relational EMR into flat feature vectors.

\subsection{EMR Data}

We are mainly interested in data on emergency department presentations
(ED) and admissions to the general hospital. The most important piece
of information is the diagnostic coding for any episodes. For ease
of exposition we assume that the diagnosis coding conforms with the
latest classification scheme, the ICD-10. Previous version or other
schemes could be also be applicable. The ICD-10 scheme is a hierarchy
of diseases covering almost all known conditions with approximately
$20,000$ codes. The codes start with a letter followed by several
digits where the digits placed later in the sequence indicate more
specific conditions. For example, injuries to the head are classified
into 10 groups, from \emph{S00} to \emph{S09}. The group \emph{S01}
means ``open wound of head'', the subgroup \emph{S011} means ``wound
in the eyelid and periocular area''. 

\begin{figure}
\begin{centering}
\includegraphics[width=0.6\textwidth]{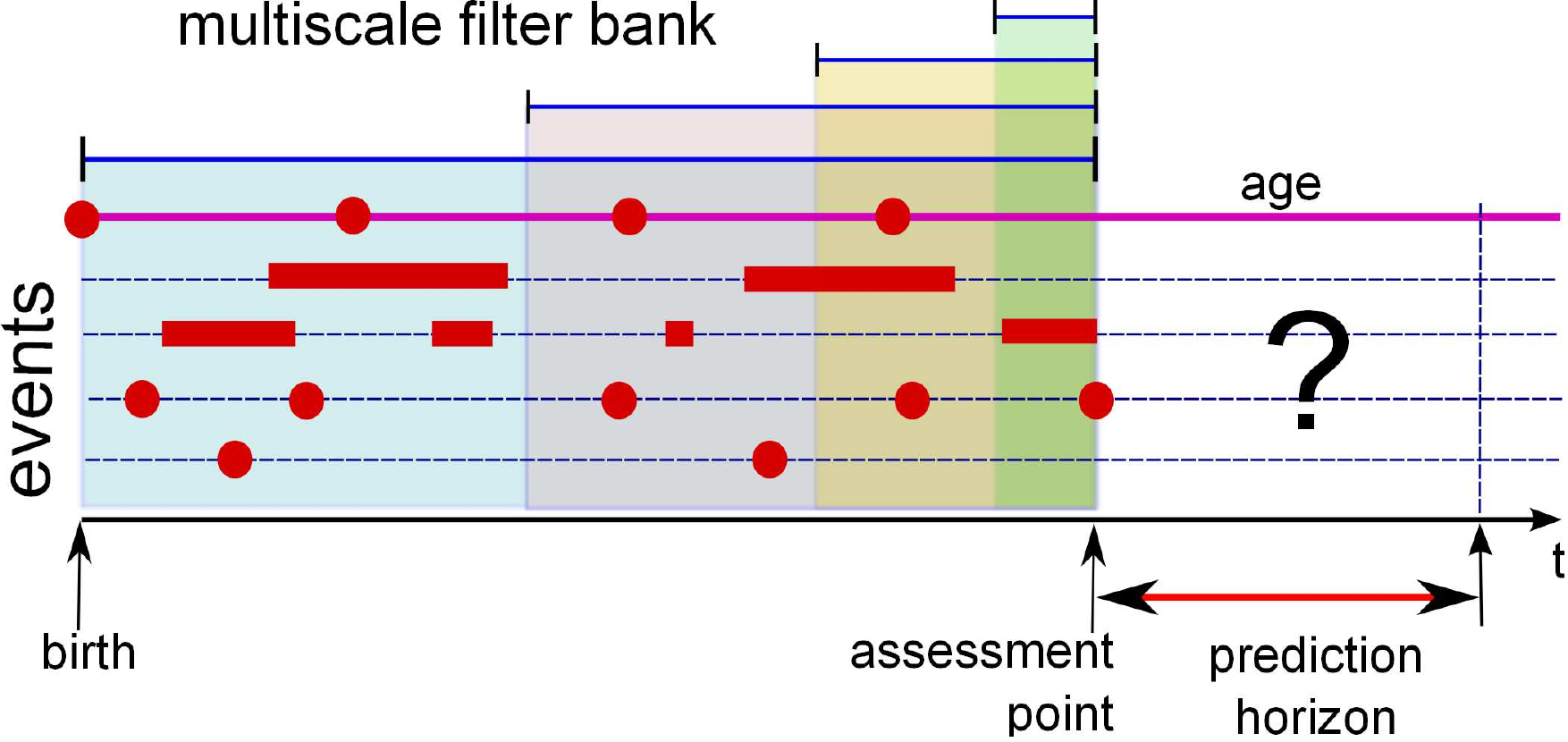}
\par\end{centering}

\caption{Clinical events represented as a temporal image, which is convoluted
with one-sided filter bank.\label{fig:Event-image-filter-bank}}
\end{figure}

In general, medical records for each patient contain time-stamped
events of different types. Thus the EMR at a given time can be represented
as a sparse 2D image (see Fig.~\ref{fig:Event-image-filter-bank}).
One dimension is time and the other dimension represents events. The
events are sparse and irregular because clinical events are often
packed into episodes. A typical episode starts with an emergency visit
followed hospitalization and ends with discharge or death. For certain
conditions such as mental health and cancers, formal risk assessments
may be performed. Emergency visits, hospitalizations and risk assessments
are major events that contain sub-events. For example, each emergency
visit includes a primary ICD-10 diagnosis code, a decision to admit,
transfer or return home. Hospitalization could be planned or come
through the emergency department. Each admission typically contains
multiple diagnoses, intervention procedures and medication prescriptions.
A risk assessment may contain a check list or ordinal ratings on multiple
risk-related items.

\subsection{Multiscale Feature Extraction using One-Sided Filter Bank \label{sec:Multiscale-Feature-Extraction}}

Our prediction problem is to stratify future risk within a window
of time using historical records. Thus at a prediction point, we transform
the EMR into a feature vector to which risk classifiers are applied.
As medical events are irregular and sparse, standard feature extraction
techniques that rely on precise timing may not be robust.  Instead,
we exploit the 2D temporal image representation using a bank of \emph{one-sided}
filters. The concept resembles filters in signal processing and vision,
except that the ``signals'' are sparse and irregular, also no future
information will be used as the filter is one-sided. 

Let $t$ be the time point of interest, $H$ be the maximum history
length. Let $v_{j}(t)$ be the observation of the event of type $j$
at time $t$, for $j=1,2..,D$. Discrete events such as diagnosis
are typically binary, i.e., $v_{j}(t)$ are the presence or absence
of a code. For continuing events such as treatment episodes, $v_{j}(t)$
is the event duration. Let $\mathcal{K}^{k}(t;\sigma_{k})$ be a kernel
function of $t$, parameterized by $\sigma_{k}$ and right-truncated
at $0$ -- that is, $\mathcal{K}^{k}(t;\sigma_{k})=0$ for $t\ge0$.
The $k$-th feature evaluated at $t$ for event $j$ is defined using
the following convolution operation:

\begin{equation}
x_{j}^{k}(t)=\sum_{h=0}^{H}\mathcal{K}^{k}(s_{k}-h;\sigma_{k})v_{j}(t-h)\label{eq:conv-feat-with-shifting}
\end{equation}
where $0\le s_{k}\le H$ denotes the delay. When $s_{k}=0$, the kernel
is effective at anytime before time $t$. In effect, the event sequence
of type $j$ is summarized throughout the history of length $H$ by
the convolution operation. However, when $s_{k}>0$, the kernel is
ineffective until $h\ge s_{k}$. This is equivalent to evaluating
the feature at $t-h$, and thus this captures the temporal progression
from $t-h$ to $t$. 

The adjustable kernel parameter $\sigma_{k}$ controls the effective
range of the kernel. This is important to differentiate acute conditions
(such as suicide ideation) from chronic conditions (such as Type I
diabetes). One useful kernel is the \emph{truncated Gaussian}
\begin{equation}
\mathcal{K}^{k}(h;\sigma_{k})=\sqrt{\frac{2}{\pi\sigma_{k}^{2}}}\exp\left(-\frac{h^{2}}{2\sigma_{k}^{2}}\right)\label{eq:Gaussian-kernel}
\end{equation}
where $\mathcal{K}^{k}(h;\sigma_{k})>0$ for $h\ge0$ and $0$ otherwise.
The hyper-parameter $\sigma_{k}$ defines the effective width of the
kernel, i.e., the response drops drastically as $h$ goes beyond $\sigma_{k}$.
The behavior is similar to the \emph{uniform kernel} with specified
width $\sigma_{k}$
\begin{equation}
\mathcal{K}^{k}(h;\sigma_{k})=\frac{1}{\sigma_{k}}\mathbf{1}\left[h\in[0,\sigma_{k}]\right]\label{eq:uniform-kernel}
\end{equation}
This kernel counts the normalized number of events falling within
a given period of time. Wavelet-like kernels could also be used to
detect the trends and recurrences.

\section{Modeling Ordinal Risk \label{sec:Risk-Modelling}}

We describe a set of ordinal regression models of risk associated
with loss functions $R(\xb_{i},y_{i};\wb)$ as in Eqs.~(\ref{eq:loss-func-lasso}).
We assume that the observed outcomes $y\in\{1,2,...,L\}$ are the
discretized version of \emph{underlying} \emph{random risks} $z\in\mathbb{R}^{m}$.
The probabilistic models are natural to estimate the probability of
a particular risk class being observed. Maximum likelihood learning
leads to the risk $R(\xb_{i},y_{i};\wb)=-\log P(y_{i}\mid x_{i};\wb)$.
Two set of classifiers are presented: classifiers with and without
shared parameters.

\subsection{Models with Shared Parameters \label{sub:Ord-Shared-Params}}

For now we assume that all the classes share the same set of parameters
$\wb$. Relaxation will be considered in the next subsection.

\subsubsection{Cumulative Classifier\label{sub:Cumulative-Models}}

This model assumes that the discrete outcomes $y$ are generated from
the \emph{one-dimensional} underlying random risk $z\in\mathbb{R}$
as follows \cite{mccullagh1980rmo}:
\begin{eqnarray*}
y & = & \begin{cases}
1 & \quad\mbox{if}\quad z\le\tau_{1}\\
l & \quad\mbox{if}\quad\tau_{l-1}<z\le\tau_{l}\\
L & \quad\mbox{otherwise}
\end{cases}
\end{eqnarray*}
where $\tau_{1}\le\tau_{2}\le...\tau_{L-1}$ are thresholds. This
essentially says that the discrete outcome is a coarse version of
the real-valued risk. The risk spectrum is the real line divided into
intervals, each of which determines the corresponding outcome. In
the form of probability distribution we have:

\begin{eqnarray*}
P(y=l\mid\xb) & = & P(\tau_{l-1}\le z\le\tau_{l}\mid\xb)\\
 & = & F(\tau_{l}\mid\xb)-F(\tau_{l-1}\mid\xb)
\end{eqnarray*}
where $F(\tau_{l}\mid\xb)$ is the cumulative distribution evaluated
at $\tau_{l}$. Choosing the form of $F(\tau_{l}\mid\xb)$ is usually
the matter of practical convenience since $x$ is unobserved and we
do not know the true underlying distribution. For example, the logistic
distribution $F(\tau_{l}\mid\xb)=\left[1+\exp\left(-(\tau_{l}-\wb^{\top}\xb)\right)\right]^{-1}$has
an interesting interpretation:
\[
\log\left(\frac{P(r\le l\mid\xb)}{P(r>l\mid\xb)}\right)=\tau_{l}-\wb^{\top}\xb
\]
i.e., the log odds at the split level $l$ is proportional to the
risk factors%
\footnote{This is known as the \emph{proportional odds model}.%
}. . The parameters to be estimated are $\wb$ and $\left\{ \tau_{l}\right\} _{l=1}^{L-1}$.

\subsubsection{Stagewise Classifier\label{sub:Stagewise-Models}}

Cumulative models assume a single risk variable that can explain the
ordinal outcomes. This assumption does not address the nature of the
risk progression -- for some patients, the risk may not reach a certain
level immediately. It may, alternatively, start from a normal condition,
and then progress upward. This suggests a stagewise model of outcomes.
The next outcome level may be attained only if the lower levels have
not been attained \cite{tutz1991sequential,Truyen:2012b}. The stagewise
process can be formalized as follows:

\begin{eqnarray*}
y & = & \begin{cases}
1\quad & \mbox{if}\quad z_{1}\le\tau_{1}\\
l\quad & \mbox{if}\quad\left\{ z_{m}\ge\tau_{m}\right\} _{m=1}^{l-1}\,\&\, z_{l}\le\tau_{l}\\
L\quad & \mbox{otherwise}
\end{cases}
\end{eqnarray*}
where $m=1,2,..,l-1$ is the index of the risk levels below the current
level $l$. Here, the transition from level $l$ to level $l+1$ is
signified by the event that the risk value passes through the level-specific
threshold $\tau_{l}$. The probability that the outcome is the lowest
is then given as:
\begin{eqnarray*}
P(y=1) & = & P(z_{1}\le\tau_{1})=F(\tau_{1})
\end{eqnarray*}
If the condition $z_{1}\le\tau_{1}$ does not hold, then we consider
level $2$, 
\begin{eqnarray*}
P(y=2\mid z\ge2) & = & P(z_{2}\le\tau_{2})=F(\tau_{2})
\end{eqnarray*}
This process continues until some level has been accepted, or we must
accept the last level $L$. Thus the probability of having the highest
level of risk, given all the lower levels have not been accepted,
is
\[
P(y=L\mid y>L-1)=1-F(\tau_{L-1})
\]
As all the decision steps rely on the same distribution $F(\tau)$,
it is natural that $\tau_{1}<\tau_{2}<...<\tau_{L-1}$.

Note that the probabilities above are \emph{conditional}. The marginal
probability of selecting a particular discrete outcome is
\[
P(y=l)=\begin{cases}
F(\tau_{1}) & \mbox{if }l=1\\
F(\tau_{l})\prod_{m=1}^{l-1}\left(1-F(\tau_{m})\right) & \mbox{if }l\in\{2,..,L-1\}\\
\prod_{m=1}^{L-1}\left(1-F(\tau_{m})\right) & \mbox{otherwise}
\end{cases}
\]

With the choice $F(\tau_{l})$ as a logistic distribution, we have
a nice interpretation
\[
\log\left(\frac{P(y=l\mid\xb)}{P(y\ge l\mid\xb)}\right)=\tau_{l}-\wb^{\top}\xb
\]
i.e., the log odds of the probability of choosing the next level,
given that all previous levels have failed, is proportional to the
risk factors $\xb$. Similar to the case of cumulative models, the
parameters to be estimated are $\wb$ and $\left\{ \tau_{l}\right\} _{l=1}^{L-1}$.

\subsection{Models with Separate Parameters \label{sub:Ord-Multi-Params}}

Models with shared parameters described in the previous subsection
treat outcome risk as one-dimensional. However, risk classes could
be qualitatively different -- for example, some people never cross
the line from an attempted suicide to a completed suicide. This suggests
treating risk classes with separate parameter sets. In general, we
have $L-1$ parameter sets $\mathbf{w}=\left\{ \wb_{1},\wb_{2},...,\wb_{L-1}\right\} $. 

Let us return to the stagewise model studied in Sec.~\ref{sub:Stagewise-Models}.
Since there are several stages, we need not assume that there is only
one underlying risk distribution. Instead, class-specific risk distribution
$F_{l}(\tau_{l};\wb_{l})$ can be used, where each class has their
own parameter $\wb_{l}$ and threshold $\tau_{l}$ for $l=1,2,...,L-1$.
The marginal distribution is then:
\[
P(y=l)=\begin{cases}
F(_{1}\tau_{1}) & \mbox{if }l=1\\
F_{l}(\tau_{l})\prod_{m=1}^{l-1}\left(1-F_{m}(\tau_{m})\right) & \mbox{if }l\in\{2,..,L-1\}\\
\prod_{m=1}^{L-1}\left(1-F_{m}(\tau_{m})\right) & \mbox{otherwise}
\end{cases}
\]

\section{Stabilizing Predictive Models \label{sec:Stabilized-Prediction}}

Under lasso-regularized training (Eq.~(\ref{eq:loss-func-lasso})),
the selected features and their weights form a model. Unfortunately,
under the sparsity constraints, models may be unstable under data
variations. The medical domain amplifies this problem even more. First,
features extracted from EMR (see Sec.~\ref{sec:Multiscale-Feature-Extraction})
are highly redundant and correlated. Lasso-based regularization, however,
tends to select only one feature between two strongly correlated ones
\cite{zou2005regularization}. Second, as features are often weakly
predictive, their selection probability is usually less than $1$.
When training data vary, this results in unstable models -- the selected
feature subset and their weights change significantly from one training
setting to another. This instability is problematic in clinical settings
because the learned model does not generalize from one cohort to another.

This section presents a remedy for this problem. First we define measures
of model stability and show how to exploit existing relational structures
in the data to stabilize the learned models.

\subsection{Stability Indices}

To quantify model stability, we assume that the models are trained
on samples drawn randomly from an unknown data distribution $\hat{\Data}\sim P(\Data)$
of the same size $n$. Each sample $b$ produces a set of features
and their weights $\wb^{b}$. Suppose that features are ranked through
some criteria $\pi$, that is we have a sequence of features $\left\{ x_{\pi(1)},x_{\pi(2)},...,x_{\pi(d)}\right\} $,
we suggest two model stability indices:
\begin{itemize}
\item \emph{Averaged selection probability (ASP) at length }$T$. This measures
how strong the features are against both selection and ranking criteria,
where $T\le d$ is the length of the selected rank list:\emph{ 
\begin{equation}
ASP@T=\frac{1}{T}\sum_{t=1}^{T}\sum_{\Data}\mathbb{I}\left(w_{\pi(t)}^{\Data}\ne0\right)P(\Data)\label{eq:ASP-index}
\end{equation}
}The term $\sum_{\Data}\mathbb{I}\left(w_{\pi(t)}^{\Data}\ne0\right)P(\Data)$
is probability that a feature is selected \cite{meinshausen2010stability}.
This index is bounded within $\left[0,1\right]$.
\item \emph{Averaged signal-to-noise ratio (SNR) at length} $T$. Assume
that the mean and standard deviation of feature weights are $\left\{ \left(\bar{w}_{j},\sigma_{j}\right)\right\} _{j=1}^{d}$.
The average SNR at $T$ is defined as:
\begin{equation}
SNR@T=\frac{1}{T}\sum_{t=1}^{T}\frac{\left|\bar{w}_{\pi(t)}\right|}{\sigma_{\pi(t)}}\label{eq:SNR-index}
\end{equation}
When no regularization is imposed, the SNR square is the well-known
Wald statistic.
\end{itemize}
In practice, since $P(\Data)$ is unknown, we propose to draw $B\gg1$
random sets from the original set $\Data$. One way is using bootstrap
\cite{efron1986bootstrap} in that each set is resampled with replacement.
Alternatively, we can subsample $50\%$ of the data \cite{meinshausen2010stability}.
The $ASP@T$ reduces to $ASP@T=\frac{1}{TB}\sum_{t=1}^{T}\sum_{b=1}^{B}\mathbb{I}\left(w_{\pi(t)}^{b}\ne0\right)$.
The $SNR@T$ stays in the same form given that mean and standard deviations
$\left\{ \left(\bar{w}_{j},\sigma_{j}\right)\right\} _{j=1}^{d}$
are estimated from the samples.

The next issue is the ranking criteria $\pi$. Selection probability
\cite{meinshausen2010stability} and the individual SNR $\frac{\left|\bar{w}_{j}\right|}{\sigma_{j}}$
could be natural criteria. Under the regression framework, one can
also employ the importance score \cite{friedman2008predictive} as:
\begin{equation}
I_{j}=\left|\bar{w}_{j}\right|\mbox{std}(x_{j})\label{eq:importance-score}
\end{equation}
where $\mbox{std}(x_{j})$ is the standard deviation of $x_{j}$.
The importance is largely scale-invariant.

\subsection{Stabilizing Sparse Models using Relational Structures \label{sub:Stablizing-Sparse-Models}}

\begin{figure*}
\begin{centering}
\begin{tabular}{cc}
\includegraphics[width=0.35\textwidth]{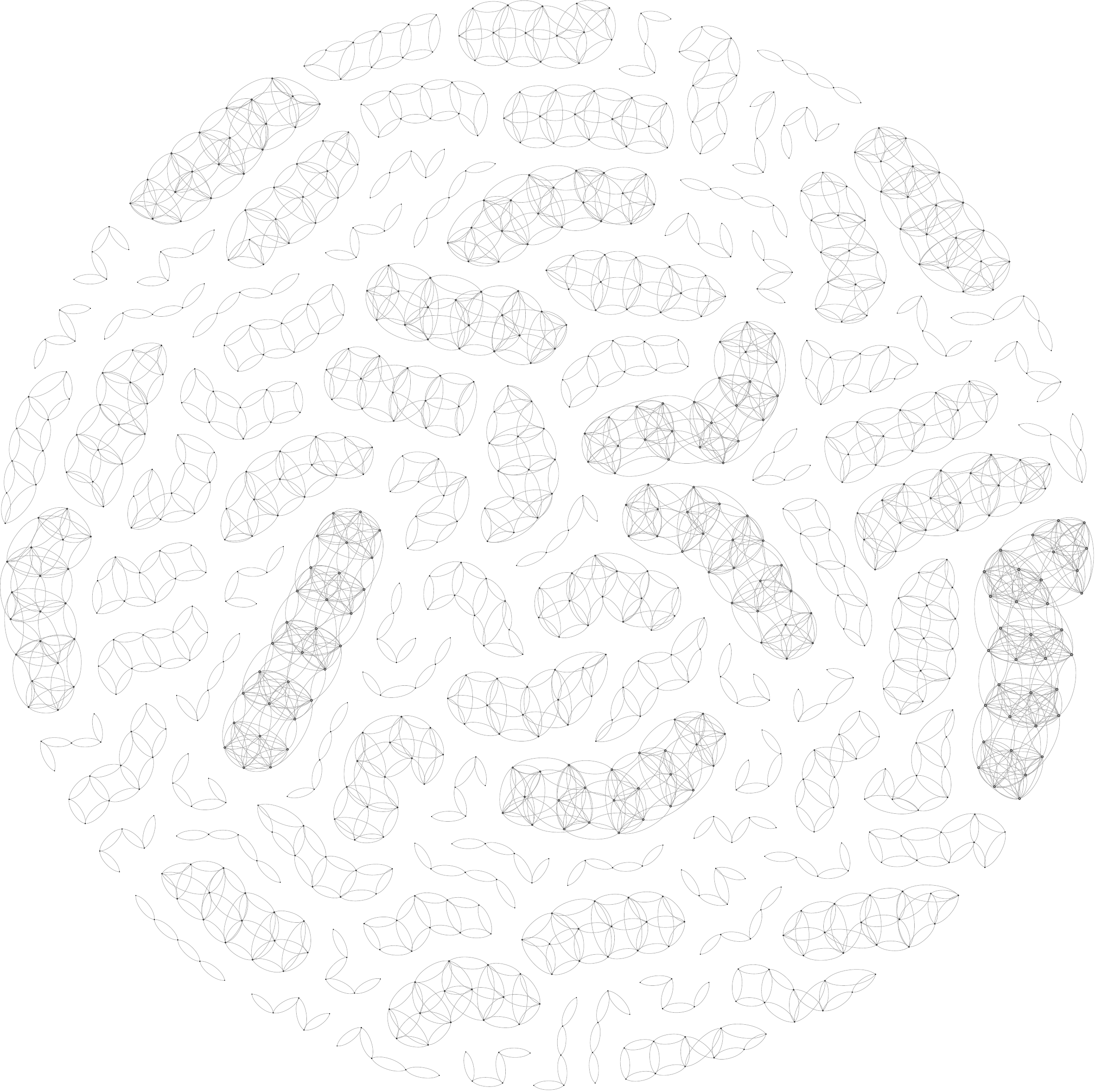} & \includegraphics[width=0.65\textwidth,height=0.35\textwidth]{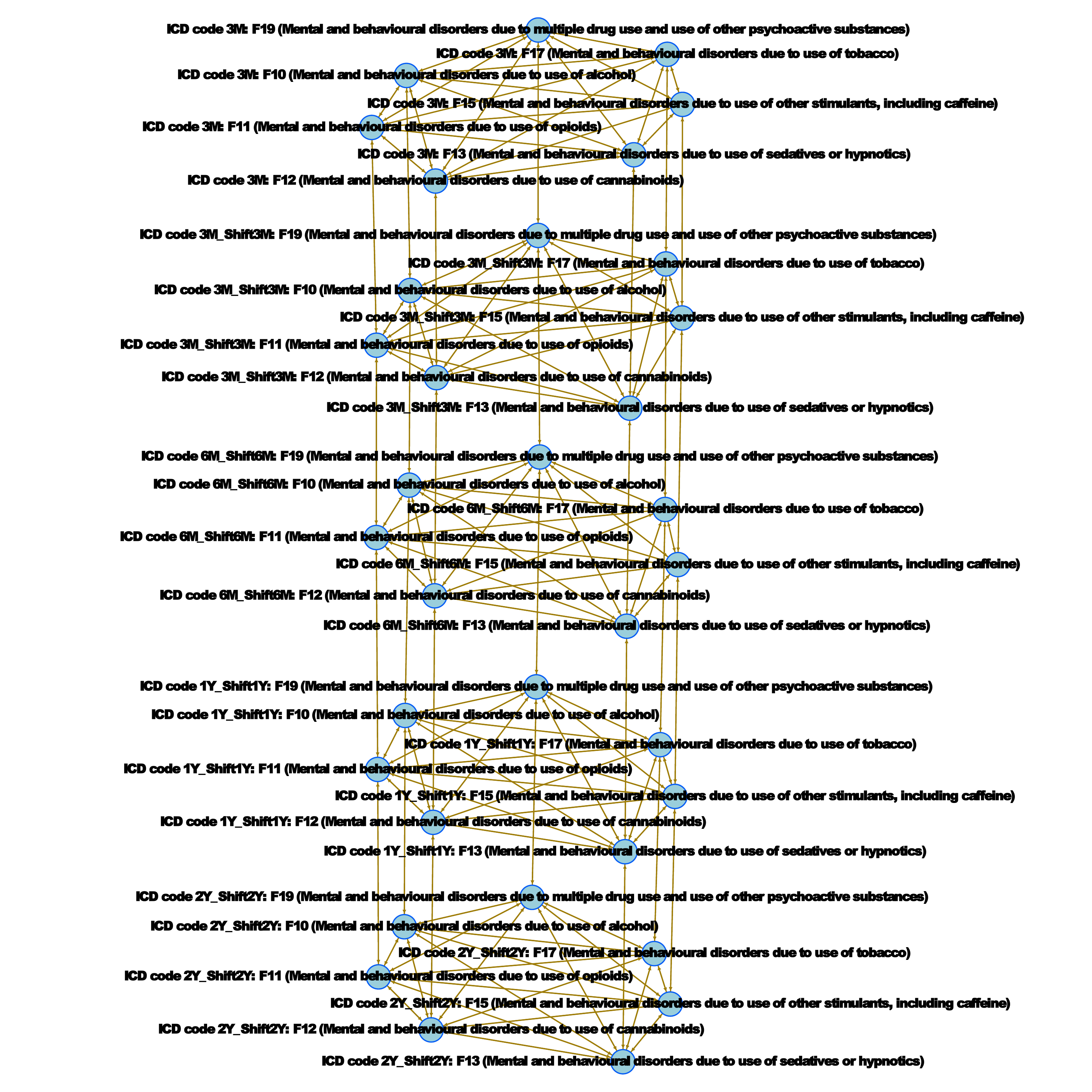}\tabularnewline
(a) ICD-10 code network & (b) Mental health sub-network\tabularnewline
\end{tabular}
\par\end{centering}

\caption{(a) Feature sub-networks for ICD-10 diagnoses from a mental health
cohort when a $2$-character sharing is counted as a link. (b) Subnet
of mental health diagnoses. Fully connected cliques are for related
codes in the same extraction period. As example, one clique in (b)
represents the group \emph{(F10-F13,F15,F17,F19}) (Mental and behavioral
disorders due to psychoactive substance use). Rare diagnoses are not
presented. See Section~\ref{sub:Data} for data description. \label{fig:Feature-sub-networks}}
\end{figure*}

To stabilize models, we exploit known structures in the features.
The first structure is temporal, wherein each event type is evaluated
at different time-scales and points, as parameterized by the kernel
width $\sigma_{k}$ and the delay $s_{k}$ in Eq.~(\ref{eq:conv-feat-with-shifting})
respectively. The other structure is the coding hierarchy, where the
code is either a diagnostic code, procedure, DRG or medication class.
The common property of the two structures is the relation among features
of the same type. For simplicity, we do not distinguish between relations
due to time-scale and those due to delay. Further, any two codes that
share the same prefix are considered to be correlated. See Fig.~\ref{fig:Feature-sub-networks}
for the sub-network of diagnostic codes used in the experiments.

Let $W\in\mathbb{R}^{n\times n}$ be the nonnegative matrix that encodes
the relation between features, i.e., $W_{ij}>0$ if features $i$
and $j$ are related and $W_{ij}=0$ otherwise. Let $S=g(W)\in\mathbb{R}^{n\times n}$
be some transform of the relation matrix into the correlation matrix.
When all classes share the same parameter set (Section~\ref{sub:Ord-Shared-Params}),
the loss function in Eq.~(\ref{eq:loss-func-lasso}) is modified
as follows:

\begin{equation}
\mbox{loss}_{2}(\wb)=\frac{1}{n}\sum_{i}R(\xb_{i},y_{i};\wb)+\alpha\left\Vert \wb\right\Vert _{1}+\beta\wb^{\top}S\wb\label{eq:loss-func-lasso-correl}
\end{equation}
where $\beta>0$ is the correlation parameter. 

When $S$ is semi-definite positive, this is equivalent to a compound
prior of a Laplace and a Gaussian of mean $\boldsymbol{0}$ and covariance
matrix of $\beta S^{-1}$. Minimizing the loss could be interpreted
as finding \emph{maximum a posterior} (MAP) solution. Let $\Omega(\wb;S)=\wb^{\top}S\wb$,
we discuss two interesting transforms from $W$ to $S$: 
\begin{itemize}
\item \emph{Linear association}. This assumes that correlation is a linear
function of relation, that is $S=D-W$, where $D$ is a positive diagonal
matrix. This translates to the regularizer:
\[
\Omega(\wb;S)=\sum_{j}D_{jj}w_{j}^{2}-\sum_{j}\sum_{k}W_{jk}w_{j}w_{k}
\]
Thus minimizing the loss tends to encourage positive correlation between
paired feature weights. The quadratic term in the left hand side is
needed to prevent the weights from going too large. One special case
is the \emph{Laplacian smoothing} \cite{li2008network,fei2010regularization},
where $D_{jj}=\sum_{k}W_{jk}$, and the above equation can be rewritten
as:
\begin{eqnarray}
\Omega(\wb;S) & = & \sum_{j}\left(\sum_{k}W_{jk}\right)w_{j}^{2}-\sum_{j}\sum_{k}W_{jk}w_{j}w_{k}\nonumber \\
 & = & \frac{1}{2}\sum_{jk}W_{jk}\left(w_{j}-w_{k}\right)^{2}\label{eq:Laplacian}
\end{eqnarray}
This regularizer treats all the relations equally.
\item \emph{Random walk}. Assume that $W$ is a probabilistic matrix, i.e.,
$\sum_{k}W_{jk}=1$ for all $j$, $W_{jk}$ is the probability of
random walk from ``state'' $j$ to state $k$. This suggests the
following regularizer \cite{sandler2008regularized}:
\begin{eqnarray}
\Omega(\wb;S) & = & \sum_{j}\left(w_{j}-\sum_{k}W_{jk}w_{k}\right)^{2}\nonumber \\
 & = & \wb^{\top}(I-W)^{\top}(I-W)\wb\label{eq:random-walk}
\end{eqnarray}
where $I$ is the identity matrix. That is, $S=(I-W)^{\top}(I-W)$,
which is symmetric nonnegative definite. This regularizer distributes
the smoothness equally among all features.
\end{itemize}
The Laplacian and random walk regularizations encourage correlated
features to have similar weights. This prevents the cases where only
one in a group of strongly correlated, predictive features is selected
by sparsity methods \cite{xu2011sparse}. The $\ell_{1}$ regularizer,
however, effectively pushes weaker feature groups toward zero weights.
The overall effect is that strong feature groups are more frequently
selected, but weak feature groups have less chance compared to the
case without network regularization. Thus the effect bears some similarity
with the sparse group methods \cite{yuan2006model}. The difference
is that our method is much more flexible with correlation structures
such as non-overlapping grouping. 

The extension to the case of class-specific parameters (Section~\ref{sub:Ord-Multi-Params})
is straightforward: 

\begin{equation}
\mbox{loss}_{3}(\mathbf{w})=\frac{1}{n}\sum_{i}R(\xb_{i},y_{i};\mathbf{w})+\sum_{l=1}^{L-1}\left(\alpha\left\Vert \wb_{l}\right\Vert _{1}+\beta\wb_{l}^{\top}S_{l}\wb_{l}\right)\label{eq:loss-func-lasso-correl-2}
\end{equation}
For simplicity we assume that $S_{l}=S$ for all $l=1,2,...,L-1$
although $S_{l}$ can encode class-specific prior knowledge (e.g.,
diabetes and hypertension are correlated under the high-risk scheme).

\section{Implementation and Results \label{sec:Implementation-and-Results}}

This section details an real-world application of the proposed framework
for suicide risk prediction. The cohort under study consists of mental
health patients who were under suicide risk assessments. Suicide stratification
has been widely acknowledged to be extremely difficult for clinicians
as there are a large number of possible risk factors but none of them
are strong enough \cite{ryan2012suicide}. This has led to recent
doubts that predictive models may not be useful at all \cite{large2011validity}.

\subsection{Data \label{sub:Data}}

\subsubsection{Mental Health Dataset}

We collected the EMR data from \textcolor{black}{Barwon Mental Health,
Drugs and Alcohol Services, the only provider in the region of 350,000
people in the central western region of Victoria in South-eastern
Australia. }For \emph{emergency attendances} (ED), there are $42K+$
recorded mental cases for $8K+$ patients in the period of 2005--2012.
For \emph{hospital admissions} (HA), there are approximately $67K$
recorded mental cases in the period of 1995--2012. The number of recorded
emergency attendances and admissions has increased over the years,
e.g., from 7,068 admissions in 2009 to 8,143 in 2010 and 8,956 in
2012. The hospitals perform \emph{suicide risk assessments} for every
mental patient under its care. The instrument has ordinal assessments
for 18 items covering all mental aspects such as suicidal ideation,
stressors, substance abuse, family support and p\textcolor{black}{sychiatric
service history. }The system recorded approximately $25K$ assessments
on $10K$ patients in the period of 2009-2012. The majority of patients
have only one assessment (62\%), followed by two assessments (17\%),
but there are about 3\% patients who have more than 10 assessments.
For those with more than one assessment, the time between two successive
assessments are: 30\% within one week, 64\% within 3 months.

We focus our study on those patients who have had a least one event
prior to a risk assessment. The dataset then has $7,578$ patients
and $17,566$ assessments. For\emph{ }each patient, we collect age,
gender, spoken language, country of birth, religion, occupation, marital
status, indigenous status, and the postcodes over time. Among patients
considered, $49.3\%$ are male and $48.7\%$ are under 35 of age at
the time of assessment.

The risk assessments are natural evaluation points for future prediction
within a given window. Future outcomes are broadly classified into
three levels of risk, based on a senior psychiatrist at Barwon Health:
class $C_{1}$ refers to low-risk outcomes, class $C_{2}$ refers
to moderate-risk (low lethality attempts), and class $C_{3}$ the
high-risk (high lethality outcomes). The classes are assigned using
a look-up table from the diagnosis codes.

The convention is that among all events occurring within the prediction
period, the class of the highest risk is chosen. For example, the
ICD-10 coded event \emph{S51} (open wound of forearm) is moderate-risk,
while\emph{ S11} (open wound of neck) would be considered as high-risk.
Typically the completed suicides are rare, and the class distributions
are imbalanced. For example, for 1-month period following the risk
assessment, there are only $24$ suicides among $409$ lethal attempts
($2.3\%$), and $834$ moderate-risk attempts ($4.8\%$). Further
class distributions are summarized in Table~\ref{tab:Outcome-class-distribution}. 

\begin{table}
\begin{centering}
\begin{tabular}{|l|r|r|r|r|}
\hline 
Horizon (day\textbf{)} & \textbf{30} & \textbf{60} & \textbf{90} & \textbf{180}\tabularnewline
\hline 
\textbf{$C_{1}$} & 16,323 & 15,750 & 15,272 & 14,291\tabularnewline
\textbf{$C_{2}$} & 834 & 1,172 & 1,436 & 19,25\tabularnewline
\textbf{$C_{3}$} & 409 & 644 & 8,58 & 1,350\tabularnewline
Suicide & 24 & 32 & 41 & 63\tabularnewline
\hline 
\end{tabular}
\par\end{centering}

\caption{Outcome class distribution following risk assessments.\label{tab:Outcome-class-distribution}}
\end{table}

\subsubsection{Data Preprocessing}

The filter bank technique (Sec.~\ref{sec:Multiscale-Feature-Extraction})
assumes that the discrete events are given. In addition to primitive
events such as emergency visits, we use several derived events. First,
ICD-10 codes and intervening procedures are mapped into their higher
level codes in their corresponding hierarchies. For example, the code
``\emph{F32.2}'' (Severe depressive episode without psychotic symptoms)
would be mapped into ``\emph{F32}'' if level 3 in the ICD-10 hierarchy
is used. This is to make the feature list robust by reducing the number
of rare codes. Second, diagnosis-related groups (DRGs) are computed
from diagnoses and interventions taking into account of disease severity
and treatment complexity. Following \cite{morris-yates2000mapping},
we derive Mental Health Diagnosis Groups (MHDGs) from ICD-10 codes
using the mapping table. The MHDGs offer an alternative view to the
mental health code groups in the ICD-10 tree. Likewise, we also map
diagnoses into 30-element comorbidities \cite{elixhauser1998comorbidity},
as they are known to be predictive of mortality/readmission risk.
From demographic data, postcode changes are tracked on the hypothetical
basis that a frequent change could signify socio-economic problems.

For robustness we only consider separate items (e.g., codes) with
more than $100$ occurrences. Other items that do not satisfy these
conditions are considered rare events. Such rare events, though statistically
less important, are critical in identifying risks if combined. We
empirically find that using diagnostic features at level $3$ in the
ICD-10 hierarchy gave the best result as they appears to balance generality
and specificity. Similarly, for intervention procedures, we use code
blocks instead of detailed codes.

The convolution operators in the filter bank (Section~\ref{sec:Multiscale-Feature-Extraction})
could be applied several times to obtain compound sum/mean statistics.
The filters, however, do not support min/max statistics. Medical risks,
on the other hand, are of great importance at the extreme, suggesting
the use of max operators. For example, out of risk items in an assessment,
an extreme risk would be sufficient to raise the alarm. Similarly,
for an item, an extreme value within the last 3 months would suggest
serious surveillance even though the current assessment is moderate.
Thus we create an extra subset of features with the max statistics,
as listed in Table~\ref{tab:Features-from-assessment}.

\begin{table}
\begin{centering}
\begin{tabular}{ll}
\hline 
(i) & Max of (overall ratings) over time\tabularnewline
(ii) & Sum of (max ratings over time) over 18 items\tabularnewline
(iii) & Sum of (mean ratings over time) over 18 items\tabularnewline
(iv) & Mean of (sum ratings over 18 items) over time\tabularnewline
(v) & Max of (sum ratings over 18 items)\tabularnewline
\hline 
\end{tabular}
\par\end{centering}

\caption{Derived features from risk assessments. Features (iii,iv) can be obtained
by applying the filters twice, one over time, the other over items.
\label{tab:Features-from-assessment}}
\end{table}

\subsection{Implementation}

\begin{table}
\begin{centering}
\begin{tabular}{|l|c|}
\hline 
\#Ordered levels & 3\tabularnewline
\#Patients & 7,578\tabularnewline
\#Data points & 17,566\tabularnewline
\#Features & 5,376\tabularnewline
\#Edges in feature network & 79,072\tabularnewline
\hline 
\end{tabular}
\par\end{centering}

\caption{Data statistics.\label{tab:Data-statistics}}

\end{table}

\subsubsection{Feature Extraction and Network Construction\label{sub:Building-Networks-of-features}}

We choose uniform kernels for ease of interpretation with the following
scale/delay pairs: $\left(\sigma_{k};s_{k}\right)\in\left\{ \left(3,0\right);\left(3,3\right);\left(6,6\right);\left(12,12\right);\left(24,24\right)\right\} $
(months), see also Eq.~(\ref{eq:conv-feat-with-shifting}). This
means that the $4$-year history is divided into non-overlapping segments:
\{$[0-3]$,$[3-6]$,$[6-12]$,$[12-24]$,$[24-48]$\} (months). The
segment size increases from the most recent to the distant past, and
this encodes the belief that old information is less specific to the
current health state. Filter responses are then normalized into the
range {[}0,1{]} and then squared.

The construction of feature network follows feature extraction. We
consider two types of network relations: \emph{Same-Code} (any two
features corresponding to the same code at different extraction periods),
and \emph{Shared-Ancestor }(any two codes that belong to the same
branch in their code hierarchy and the same extraction period). There
are four coding types: diagnoses, intervention procedures, MHDGs,
DRGs, and medications. For each coding type, the first two characters
and digits are used to identify the \emph{Shared-Ancestor} relation.
For example, diagnosis code \emph{F31} (Bipolar affective disorder)
and \emph{F32} (Depressive episode) would be linked, but \emph{F31}
and \emph{F20} (Schizophrenia) would not. The resulting network has
$5.4K$ nodes and $97.1K$ edges. Fig.~\ref{fig:Feature-sub-networks}(a)
shows the entire ICD-10 network (less the rare diagnoses), and Fig.~\ref{fig:Feature-sub-networks}(b)
displays the sub-network corresponding to mental health diagnoses.
Table~\ref{tab:Data-statistics} summarizes the statistics of the
data.

\subsubsection{Learning Classifiers}

For cumulative and stagewise classifiers (Sec.~\ref{sub:Cumulative-Models}
and Sec.~\ref{sub:Stagewise-Models}), logistic distributions for
the underlying random risks are used. We approximate the $\ell_{1}$-norm
$|x|$ in Eqs.~(\ref{eq:loss-func-lasso},\ref{eq:loss-func-lasso-correl},\ref{eq:loss-func-lasso-correl-2})
by the Huber-like loss function, where $H(w)=0.5w^{2}/\epsilon$ if
$|w|\le\epsilon$ and $H(w)=|w|-0.5\epsilon$ otherwise for some small
$\epsilon>0$. This loss function behaves like $|w|$ when $|w|$
is large compared to $\epsilon$. However, the gradient is smooth:
$H'(w)=w/\epsilon$ if $|w|\le\epsilon$ and $H'(w)=\mbox{sign}(w)$
otherwise. This makes it possible to use fast large-scale optimization
algorithms such as L-BFGS. Once the optimization has converged, features
are selected if their absolute weights are larger than $10^{-3}$.

\subsubsection{Evaluation Protocol}

We use $10$-fold cross-validation \emph{in the patient space}, that
is, the set of unique patients is divided in to subsets of equal size.
Classifiers are trained on data for $9$ subsets and tested on the
remaining subset. The results are reported for all validation subsets
combined. Note that this can be a stronger test than the cross-validation
in the data space because it removes any potential patient-specific
correlation (also known as \emph{random-effects}). We employ several
performance measures: For each outcome class, we use \emph{recall}
$R$ -- the portion of groundtruth class that is correctly identified;
the \emph{precision} $P$ -- the portion of identified class that
is actually correct; and the \emph{F-score} -- their harmonic mean
$F_{1}=2RP/(R+P)$. While these measures are appropriate for class-specific
performance, they do not represent misclassification in the ordinal
setting well. For that reason, we also use \emph{Macro-averaged} \emph{Mean
Average Error} (Macro-MAE) \cite{baccianella2009evaluation} -- the
discrepancy between the true and the predicted risk levels, adjusted
for data imbalance.

\subsection{Risk Prediction}

\subsubsection{Sensitivity to Hyperparameters}

\begin{figure}
\begin{centering}
\begin{tabular}{cc}
\includegraphics[width=0.45\columnwidth]{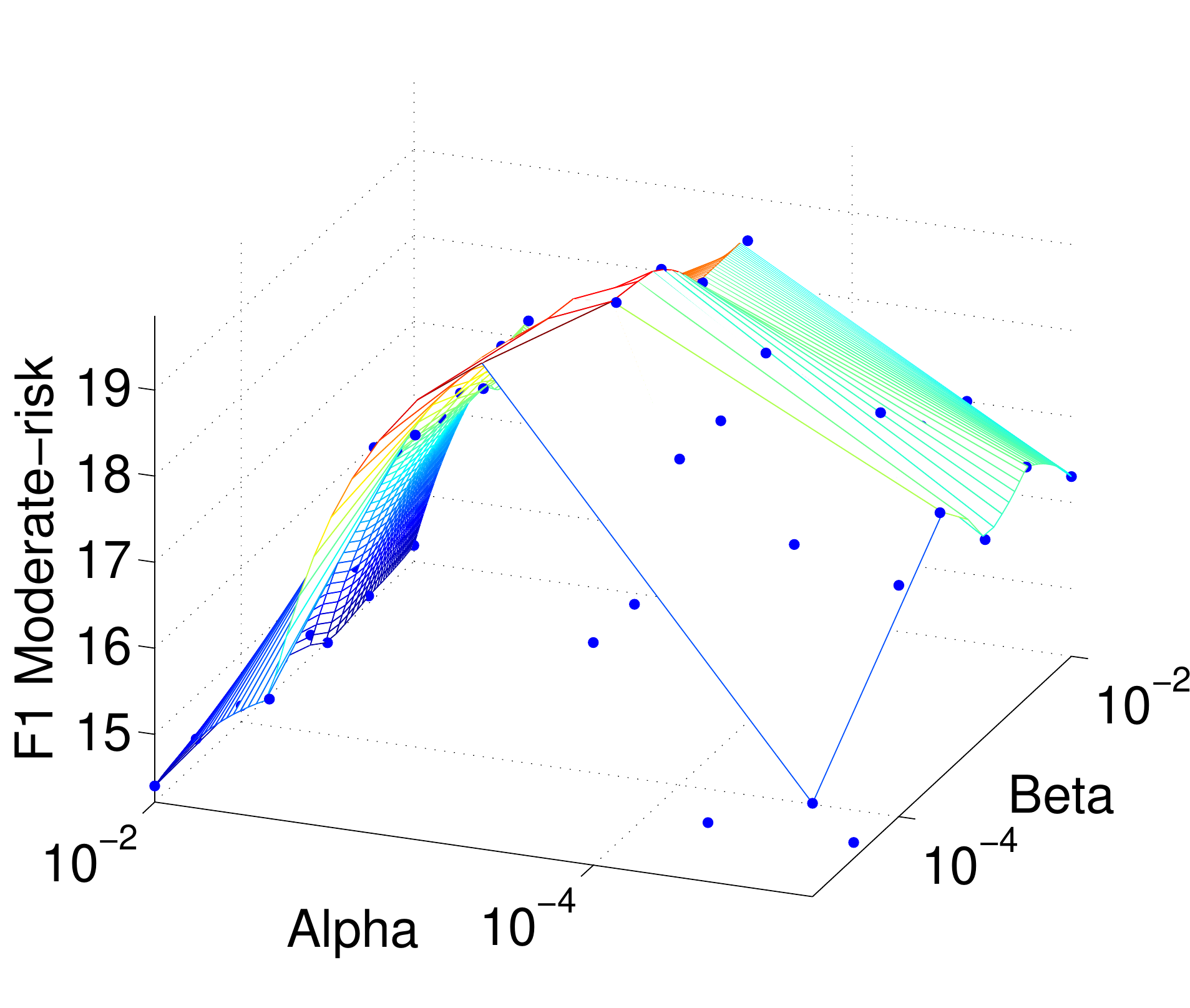} & \includegraphics[width=0.45\columnwidth]{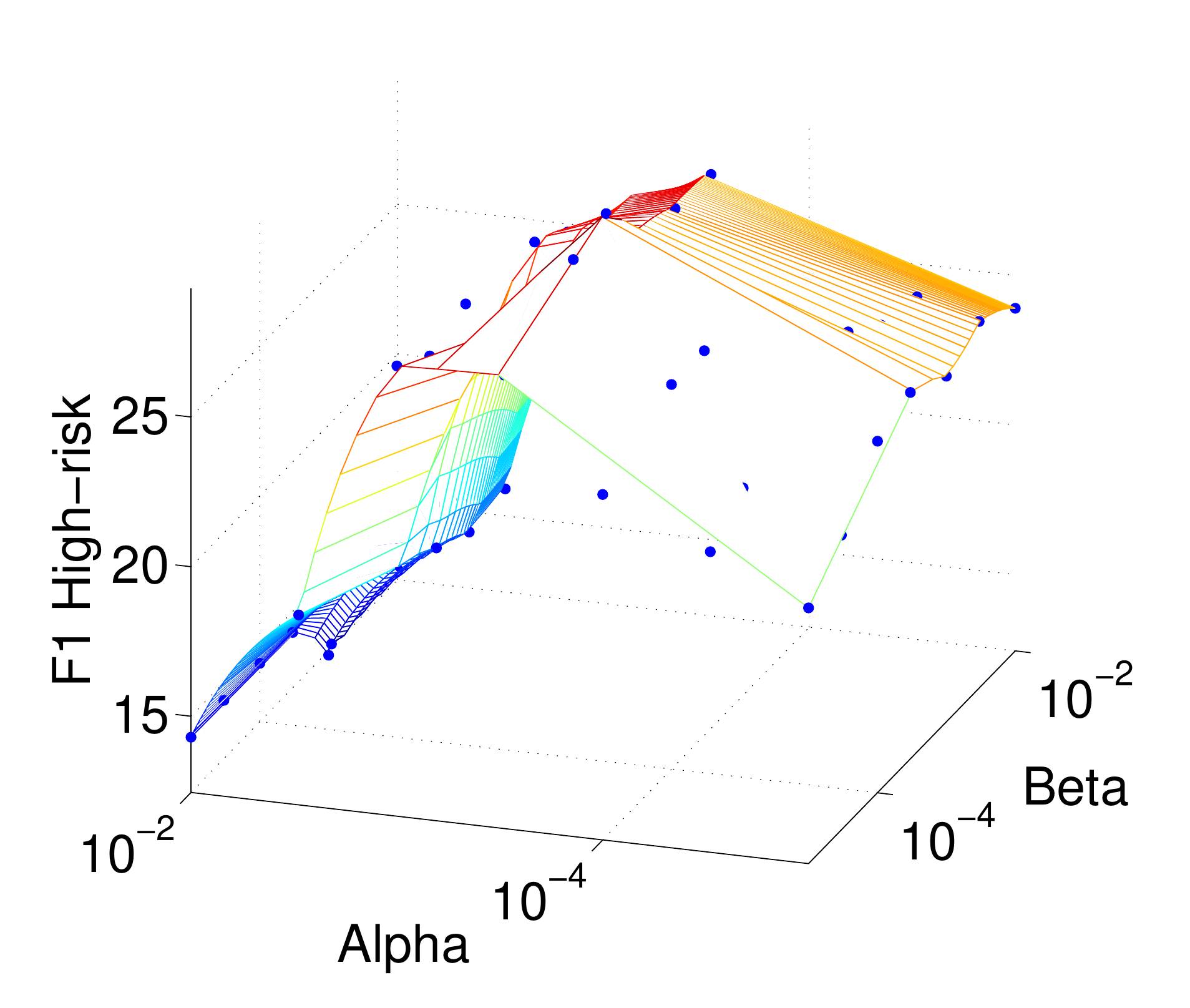}\tabularnewline
(a) Moderate-risk & (b) High-risk\tabularnewline
\end{tabular}
\par\end{centering}

\caption{Performance ($F$-scores) of the cumulative classifier (Section~\ref{sub:Cumulative-Models})
against hyperparameters in Eq.~(\ref{eq:loss-func-lasso-correl}):
$\alpha$ (sparsity) and $\beta$ (stability, using Laplacian methods,
Eq.~(\ref{eq:Laplacian})). $\beta=0$ reduces to standard lasso-based
sparse models. Similar behaviors are also observed with other classifiers.
\label{fig:Performance-versus-hyperparameters}}
\end{figure}

There are two hyperparameters in our objective function in Eq.~(\ref{eq:loss-func-lasso-correl}):
the $\ell_{1}$-norm regularization factor $\alpha$ and the network
regularization factor $\beta$. These two factors serve different
purposes: The $\ell_{1}$-norm regularization as an embedded feature
selection mechanism, and the network regularization for stabilizing
the models. To investigate the sensitivity of the final performance
against these hyperparameters, we perform a grid search in the set
\{$10^{-5},$$3\times10^{-5}$,$10^{-4}$,$3\times10^{-4}$,$10^{-3}$,$3\times10^{-3}$,$10^{-2}$\}
for each. Figs.~\ref{fig:Performance-versus-hyperparameters}(a,b)
report the $F_{1}$-score measures for the moderate-risk ($C_{2}$
class) and high-risk ($C_{3}$ class) outcomes within $3$ months
under cumulative classifiers (Section~\ref{sub:Cumulative-Models}).
The $F_{1}$-scores in both risk classes critically depends on $\alpha$
but are relatively stable against $\beta$. The former dependency
is expected: large $\alpha$ generally leads to sparser models, and
thus less overfitting. When the sparsity reaches the right level --
at $\alpha=3\times10^{-4}$ -- the predictive power peaks. The small
effect of $\beta$ on the performance is interesting but not surprising.
As large $\beta$ forces linked features to have similar weights,
the feature influence is rearranged but overall their total effect
remains largely unchanged. Thus in what follows, unless specified
otherwise, we fix the sparsity hyperparameter as $\alpha=3\times10^{-4}$
for all classifiers.

\subsubsection{Comparison Against Clinicians}

\begin{table}
\begin{centering}
\begin{tabular}{|l|c|cccc|cccc|}
\cline{3-10} 
\multicolumn{1}{l}{} &  & \multicolumn{4}{c|}{\textcolor{blue}{$C_{2}$}} & \multicolumn{4}{c|}{\textcolor{red}{$C_{3}$}}\tabularnewline
\cline{2-10} 
\multicolumn{1}{l|}{} & \emph{Macro-MAE} & \textcolor{blue}{Cases} & \textcolor{blue}{$R$} & \textcolor{blue}{$P$} & \textcolor{blue}{\emph{$F_{1}$}} & \textcolor{red}{Cases} & \textcolor{red}{$R$} & \textcolor{red}{$P$} & \textcolor{red}{$F_{1}$}\tabularnewline
\hline 
\emph{Clinician} & 0.826 & \textcolor{blue}{338} & \textcolor{blue}{23.5 } & \textcolor{blue}{11.7 } & \textcolor{blue}{15.6} & \textcolor{red}{70} & \textcolor{red}{8.2} & \textcolor{red}{12.9} & \textcolor{red}{{} 10.0}\tabularnewline
\hline 
CUMUL & 0.675 & \textcolor{blue}{429} & \textcolor{blue}{29.9} & \textcolor{blue}{14.8} & \textcolor{blue}{19.8} & \textcolor{red}{282} & \textcolor{red}{32.9} & \textcolor{red}{26.0} & \textcolor{red}{29.0}\tabularnewline
STW (Shared) & 0.681 & \textcolor{blue}{417} & \textcolor{blue}{29.0} & \textcolor{blue}{14.4} & \textcolor{blue}{19.3} & \textbf{\textcolor{red}{263}} & \textcolor{red}{30.7} & \textcolor{red}{25.6} & \textcolor{red}{27.9}\tabularnewline
STW (Multi) & \textbf{0.672} & \textcolor{blue}{418} & \textcolor{blue}{29.1} & \textcolor{blue}{14.5} & \textcolor{blue}{19.3} & \textcolor{red}{289} & \textbf{\textcolor{red}{33.7}} & \textcolor{red}{26.7} & \textbf{\textcolor{red}{29.8}}\tabularnewline
\hline 
\end{tabular}
\par\end{centering}

\caption{Predicting 3-month risk on EMR data \emph{without} model stabilization
(standard sparse models). $C_{2}$ = moderate-risk, $C_{3}$ = high-risk,
$R$ = Recall, $P$ = Precision, in percentages\emph{. CUMUL = }Cumulative
model\emph{, STW = }Stagewise model. \label{tab:Performance-of-classifiers} }
\end{table}

We first evaluate the predictive power of the mandatory risk assessments
being performed by Barwon Health. Using the overall assessment (risk
ratings of $3$ and $4$ are high-risk, $2$ moderate-risk, and ratings
of $1$ and $0$ are low-risk), the performance on the high-risk class
for $3$ month horizons is quite poor: $R=8.2\%$, $P=12.9\%$, $F_{1}=10.0\%$.
There are $14$ suicide cases ($34\%$) detected from the $C_{2}$
and $C_{3}$ assignments. Tab.~\ref{tab:Performance-of-classifiers}
lists more details. Machine learning algorithms applied to EMR data
significantly outperform the mental health professionals to a large
margin. For moderate-risk prediction, the $F_{1}$-score by machines
reach roughly $19.5\%$, which are $25\%$ improvement over the score
by clinicians. The differentials are even better for the high-risk
class: the improvement are more than $180\%$. When accounting for
class imbalance (Table~\ref{tab:Outcome-class-distribution}), machine
learning models win by roughly $18\%$ on the Macro-averaged MAE measure.

\begin{table}
\begin{centering}
\begin{tabular}{|l|c|c|c|}
\cline{2-4} 
\multicolumn{1}{l|}{} & \textbf{\textcolor{red}{Suicide}} & \textbf{\textcolor{blue}{Resource}} & \textbf{FN}\tabularnewline
\multicolumn{1}{l|}{} & \textcolor{red}{(out of 41)} & \textbf{\textcolor{blue}{cost}}\textcolor{blue}{{} ($\uparrow\%$)} & ($\downarrow\%$)\tabularnewline
\hline 
\emph{Clinician} & \textcolor{red}{14} & \textcolor{blue}{3,444 (0.0)} & 1,530 (0.0)\tabularnewline
\hline 
CUMUL & \textcolor{red}{30} & \textcolor{blue}{3,821 (10.1)} & 1,309 (29.5)\tabularnewline
STW (Shared) & \textcolor{red}{29} & \textcolor{blue}{3,920 (13.8)} & 1,138 (38.8)\tabularnewline
STW (Multi) & \textcolor{red}{29} & \textcolor{blue}{3,973 (15.4)} & 1,108 (40.4)\tabularnewline
\hline 
\end{tabular}
\par\end{centering}

\caption{Predicting 3-month risk on EMR data \emph{without} model stabilization
(standard sparse models). \emph{CUMUL = }Cumulative model\emph{, STW
= }Stagewise model. Resource cost is the total number of cases assigned
as moderate/high-risk. \emph{FN} = false negatives, which are the
risky cases wrongly classified as low-risk. The symbols $\uparrow$
and $\downarrow$ denote the amount increase or decrease relative
to the reference figures by clinicians. \label{tab:Cost-FN} }
\end{table}

The practical significance of the difference is remarkable. Assuming
for simplicity that the management cost, on average, is similar for
both the moderate and high risk classes. Thus a detection of moderate
or high risk costs one basic resource unit. The machine algorithms
typically use slightly more resource units than clinicians but with
less false negatives (Table~\ref{tab:Cost-FN}). For example, the
stagewise classifier with shared parameters (Sec.~\ref{sub:Stagewise-Models})
leads to $3,920$ resource units ($13.8\%$ higher than those by clinicians),
but with $1,138$ false negatives ($25.6\%$ lower than those by clinicians).
The significance may be amplified considering that the social cost
for false negatives is much more serious than hospital resources.
In terms of suicide detection, the machine detects $29-30$ cases,
which are more than double the number detected by human ($14$ cases). 

Next we examine whether using machine learning can improve the prediction
using the risk assessments itself. We ran all classifiers on both
assessment-based features, and EMR-based features using the Laplacian
stabilization. The prediction horizons were 1,2, 3 or 6 months. As
reported in Tables~\ref{tab:perforamance-versus-time-F1}~and~\ref{tab:perforamance-versus-time-Macro-MAE},
machine learning methods trained on risk assessments consistently
outperform clinicians. The results also demonstrate that using EMR
alone is even better. This is significant because EMRs already exist
in the data warehouse, that is, we can make predict without any extra
cost.

\begin{table}
\begin{centering}
\begin{tabular}{|lc|cccc|cccc|}
\hline 
 &  & \multicolumn{4}{c|}{\textbf{\emph{$C_{2}$}}} & \multicolumn{4}{c|}{\textbf{\emph{$C_{3}$}}}\tabularnewline
\cline{3-10} 
\textbf{Classifier} & \textbf{Data} & \emph{1mm} & \emph{2mm} & \emph{3mm} & \emph{6mm} & \emph{1mm} & \emph{2mm} & \emph{3mm} & \emph{6mm}\tabularnewline
\hline 
\textcolor{magenta}{\emph{Clinician}} & \textcolor{magenta}{\emph{RA}} & \textcolor{magenta}{\emph{11.9}} & \textcolor{magenta}{\emph{14.7}} & \textcolor{magenta}{\emph{15.6}} & \textcolor{magenta}{\emph{15.7}} & \textcolor{magenta}{\emph{9.0}} & \textcolor{magenta}{\emph{9.1}} & \textcolor{magenta}{\emph{10.0}} & \textcolor{magenta}{\emph{10.0}}\tabularnewline
\hline 
\multirow{2}{*}{CUMUL} & \textcolor{blue}{\emph{RA}} & \textcolor{blue}{13.1} & \textcolor{blue}{16.0} & \textcolor{blue}{17.4} & \textcolor{blue}{19.3} & \textcolor{blue}{13.3} & \textcolor{blue}{16.2} & \textcolor{blue}{20.0} & \textcolor{blue}{25.3}\tabularnewline
 & \emph{EMR} & 13.8 & 17.7 & 19.0 & 20.1 & 16.5 & 22.1 & 27.9 & 27.7\tabularnewline
\hline 
\multirow{2}{*}{STW (Shared} & \textcolor{blue}{\emph{RA}} & \textcolor{blue}{11.9} & \textcolor{blue}{15.68} & \textcolor{blue}{16.9} & \textcolor{blue}{19.0} & \textcolor{blue}{13.6} & \textcolor{blue}{18.3} & \textcolor{blue}{22.2} & \textcolor{blue}{27.2}\tabularnewline
 & \emph{EMR} & 13.8 & \textbf{18.4} & 19.2 & 19.9 & \textbf{16.7} & \textbf{25.4} & \textbf{29.4} & \textbf{30.9}\tabularnewline
\hline 
\multirow{2}{*}{STW (Multi)} & \textcolor{blue}{\emph{RA}} & \textcolor{blue}{13.0} & \textcolor{blue}{16.1} & \textcolor{blue}{17.1} & \textcolor{blue}{19.6} & \textcolor{blue}{14.2} & \textcolor{blue}{17.1} & \textcolor{blue}{21.9} & \textcolor{blue}{26.7}\tabularnewline
 & \emph{EMR} & \textbf{14.0} & 17.4 & \textbf{19.5} & \textbf{20.6} & \textbf{16.7} & 22.5 & 28.4 & 29.7\tabularnewline
\hline 
\end{tabular}
\par\end{centering}

\caption{$F_{1}$-scores (\%) at different prediction horizons (1,2,3,6 months).
$C_{2}$ = moderate-risk, $C_{3}$ = high-risk, \emph{CUMUL = }Cumulative
model\emph{, STW = }Stagewise model\emph{, RA} = Risk assessments,
\emph{EMR} = Electronic Medical Record. Laplacian stabilization was
used, $\alpha=3\times10^{-3}$. \label{tab:perforamance-versus-time-F1}}

\end{table}

\begin{table}
\begin{centering}
\begin{tabular}{|lc|cccc|}
\hline 
\textbf{Classifier} & \textbf{Data} & \emph{1mm} & \emph{2mm} & \emph{3mm} & \emph{6mm}\tabularnewline
\hline 
\textcolor{magenta}{\emph{Clinician}} & \textcolor{magenta}{\emph{RA}} & 0.786 & 0.811 & 0.826 & 0.852\tabularnewline
\hline 
\multirow{2}{*}{CUMUL} & \textcolor{blue}{\emph{RA}} & 0.727 & 0.735 & 0.735 & 0.745\tabularnewline
 & \emph{EMR} & 0.688 & 0.677 & 0.675 & 0.712\tabularnewline
\hline 
\multirow{2}{*}{STW (Shared} & \textcolor{blue}{\emph{RA}} & 0.722 & 0.720 & 0.724 & 0.737\tabularnewline
 & \emph{EMR} & 0.671 & \textbf{0.655} & 0.681 & \textbf{0.704}\tabularnewline
\hline 
\multirow{2}{*}{STW (Multi)} & \textcolor{blue}{\emph{RA}} & 0.721 & 0.726 & 0.725 & 0.740\tabularnewline
 & \emph{EMR} & \textbf{0.670} & 0.682 & \textbf{0.672} & 0.705\tabularnewline
\hline 
\end{tabular} 
\par\end{centering}

\caption{Macro-MAE at different prediction horizons (1,2,3,6 months). \emph{CUMUL
= }Cumulative model\emph{, STW = }Stagewise model\emph{, RA} = Risk
assessments, \emph{EMR} = Electronic Medical Record. Laplacian stabilization
was used, $\alpha=3\times10^{-3}$. \label{tab:perforamance-versus-time-Macro-MAE}}
 
\end{table}

\subsection{Model Stability}

\begin{figure}
\begin{centering}
\begin{tabular}{cc}
\includegraphics[width=0.4\textwidth]{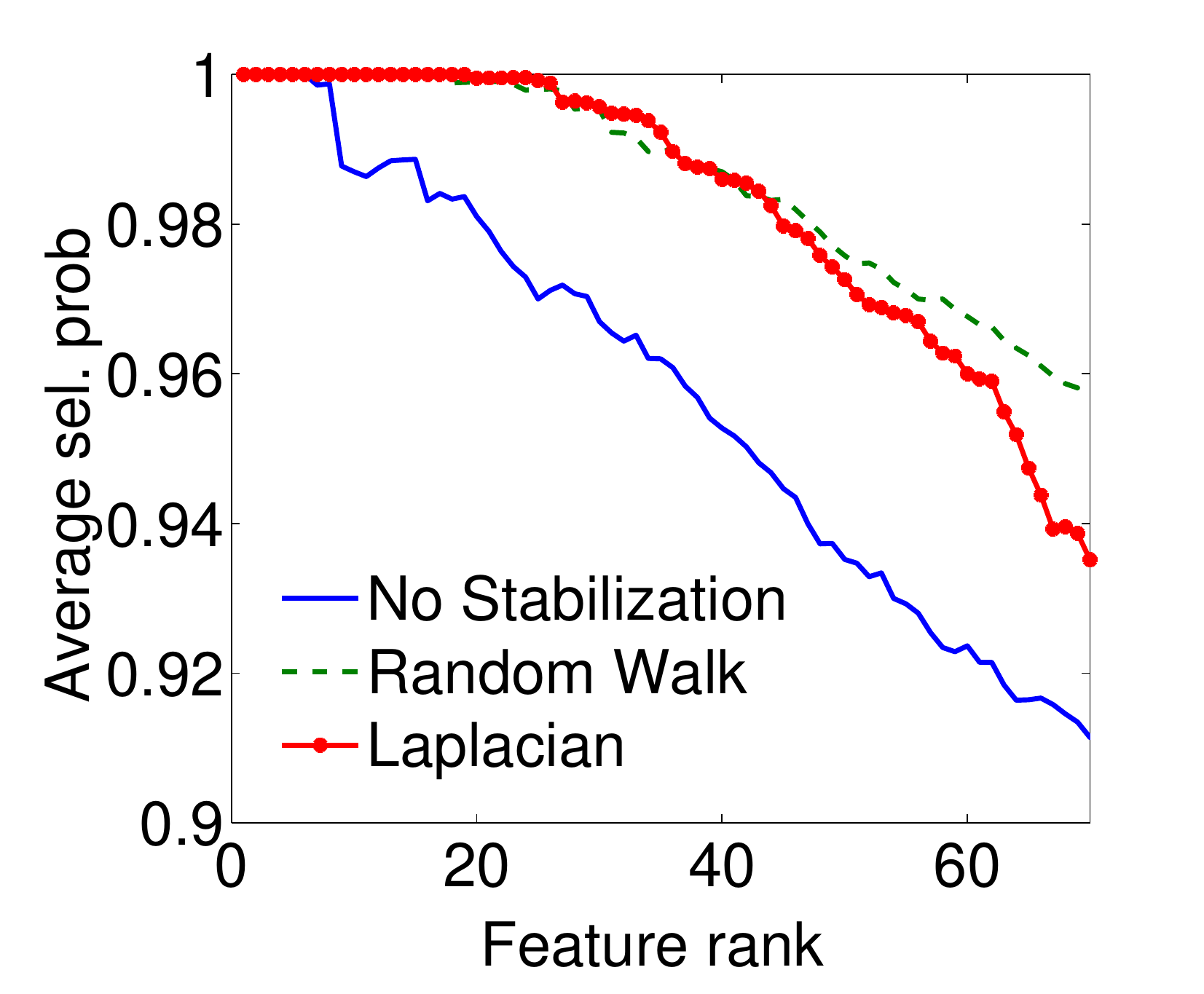} & \includegraphics[width=0.4\textwidth]{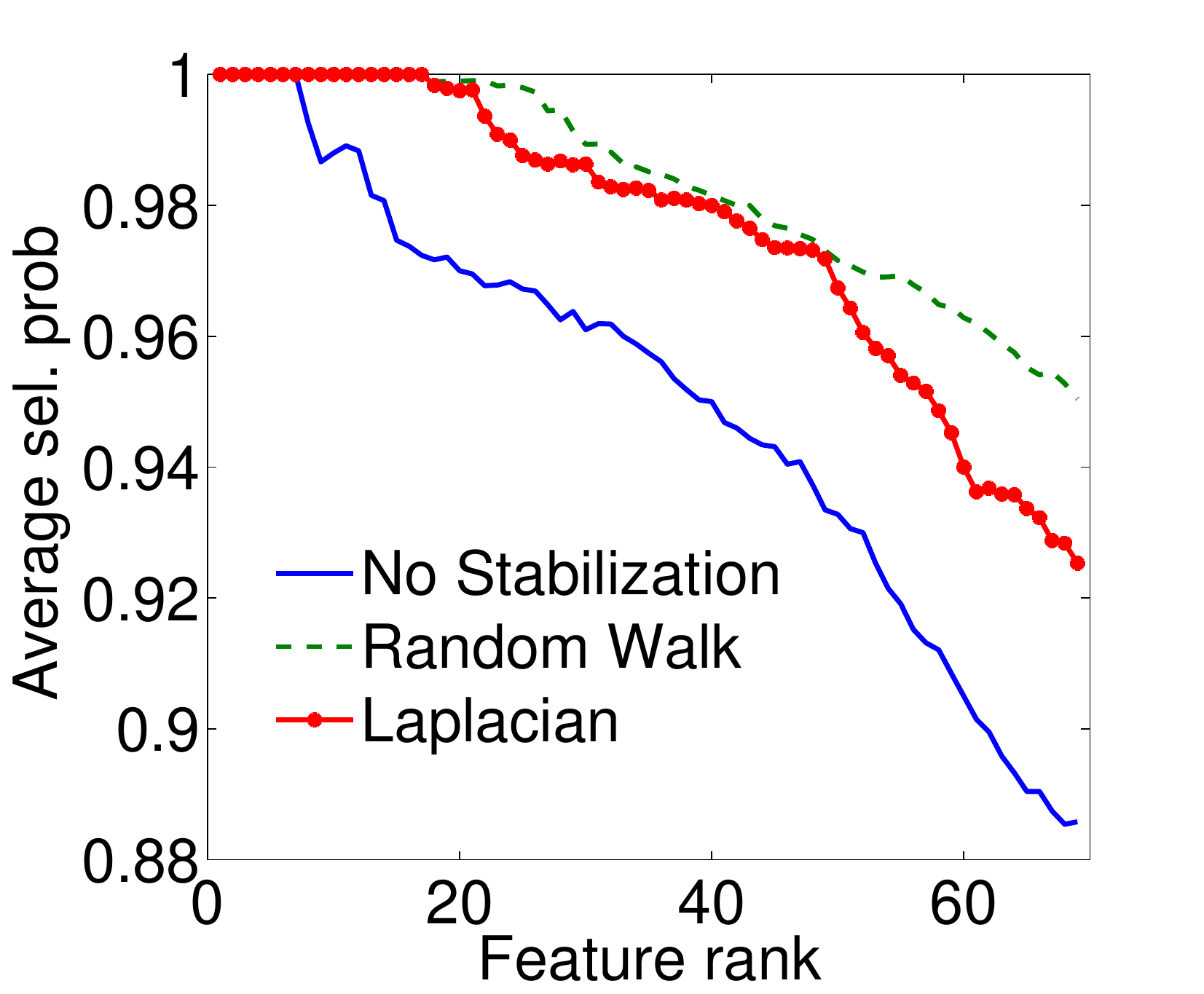}\tabularnewline
(a) Cumulative, $C_{2}+C_{3}$ & (b) Stagewise (Shared), $C_{2}+C_{3}$\tabularnewline
\includegraphics[width=0.4\textwidth]{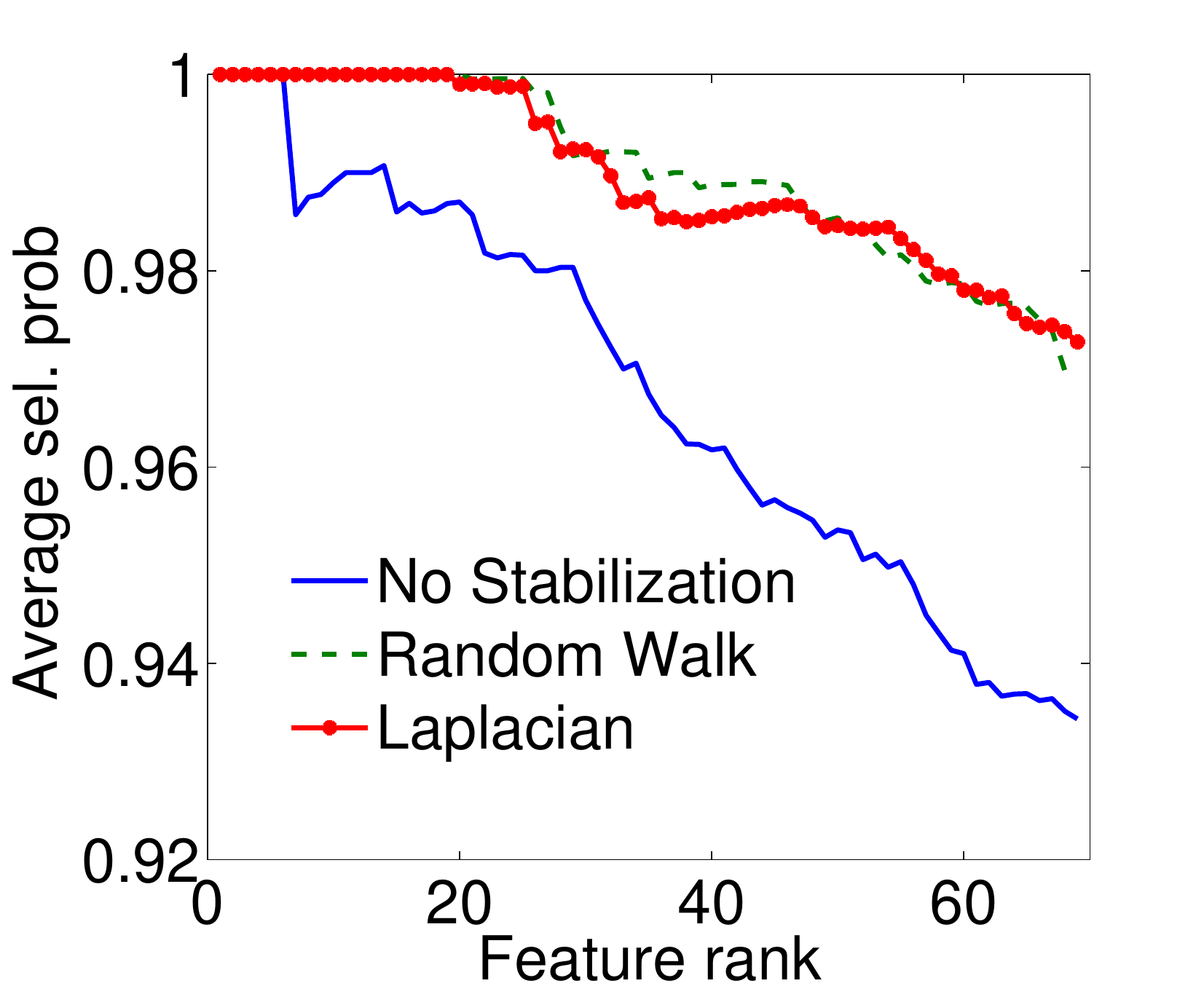} & \includegraphics[width=0.4\textwidth]{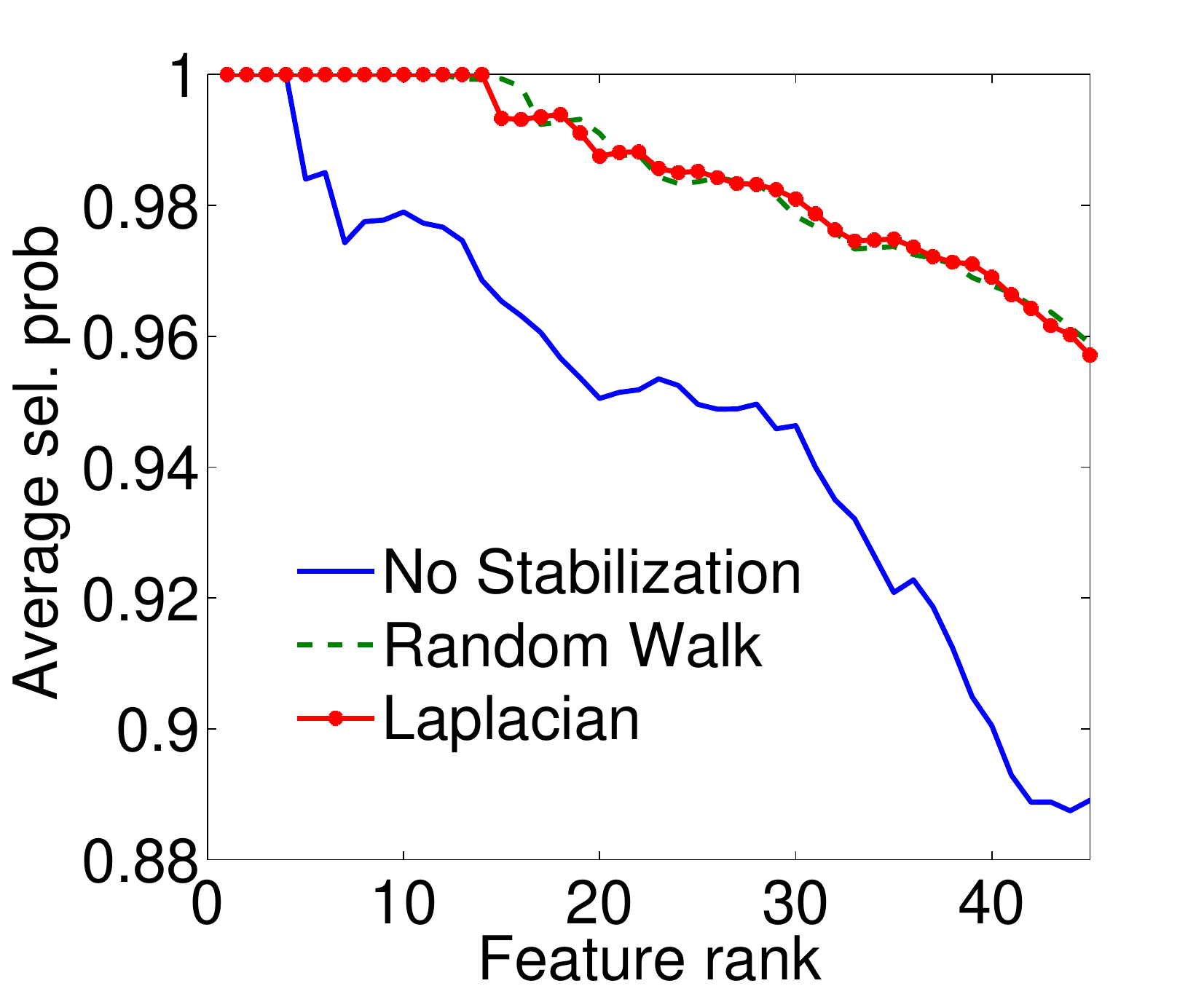}\tabularnewline
(c) Stagewise (Separate), $C_{2}$ & (d) Stagewise (Separate), $C_{3}$\tabularnewline
\end{tabular}
\par\end{centering}

\caption{Average selection probability ($ASP@T$ -- see Eq.~(\ref{eq:ASP-index})),
evaluated at different rank list sizes (the larger probability, the
more stable models). ``No Stabilization'' means the standard lasso
framework with $\beta=0$. For others, $\alpha=3\times10^{-4},\beta=3\times10^{-3}$.\label{fig:Average-selection-probability}}
\end{figure}

\begin{figure}
\begin{centering}
\begin{tabular}{cc}
\includegraphics[width=0.4\textwidth]{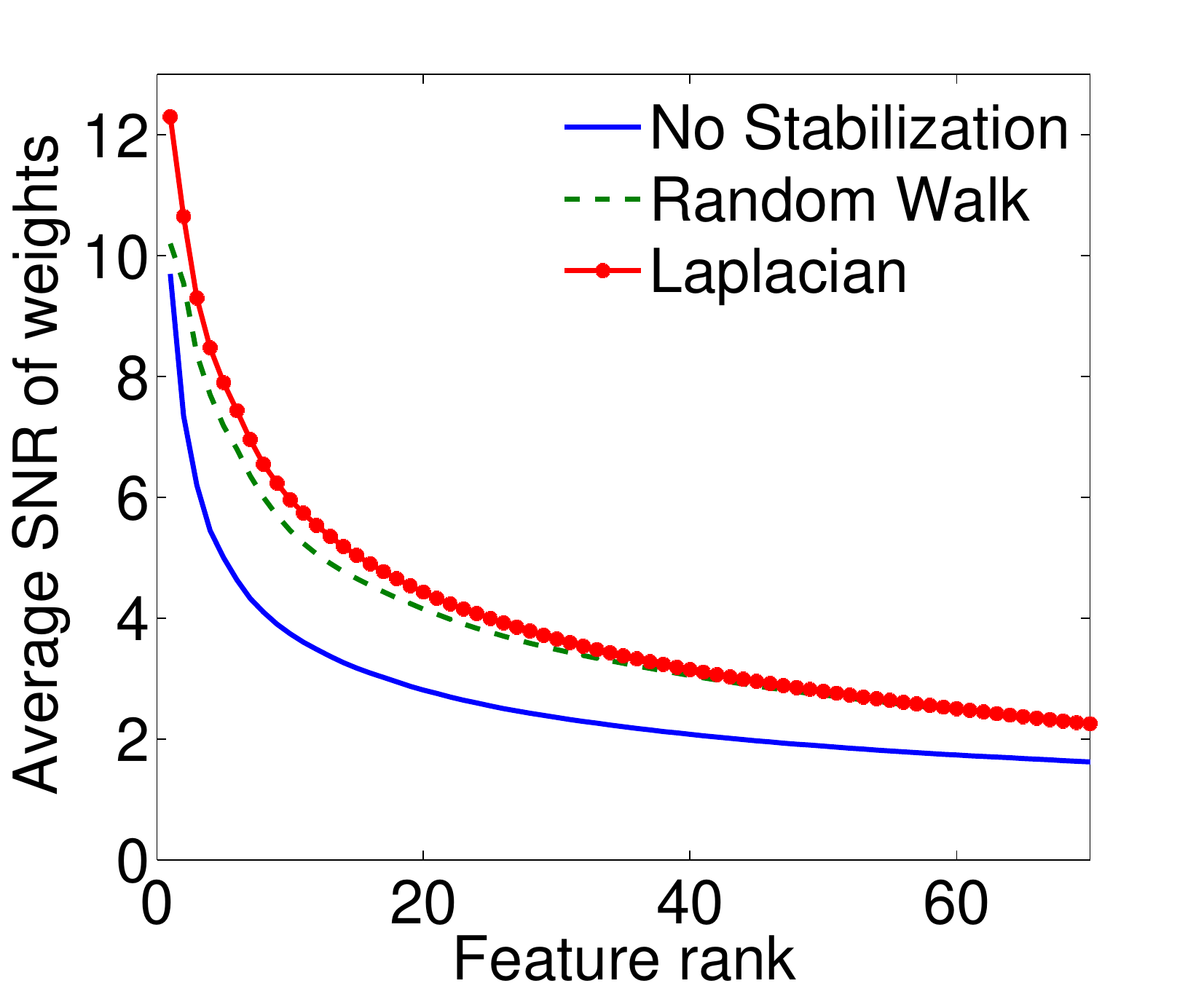} & \includegraphics[width=0.4\textwidth]{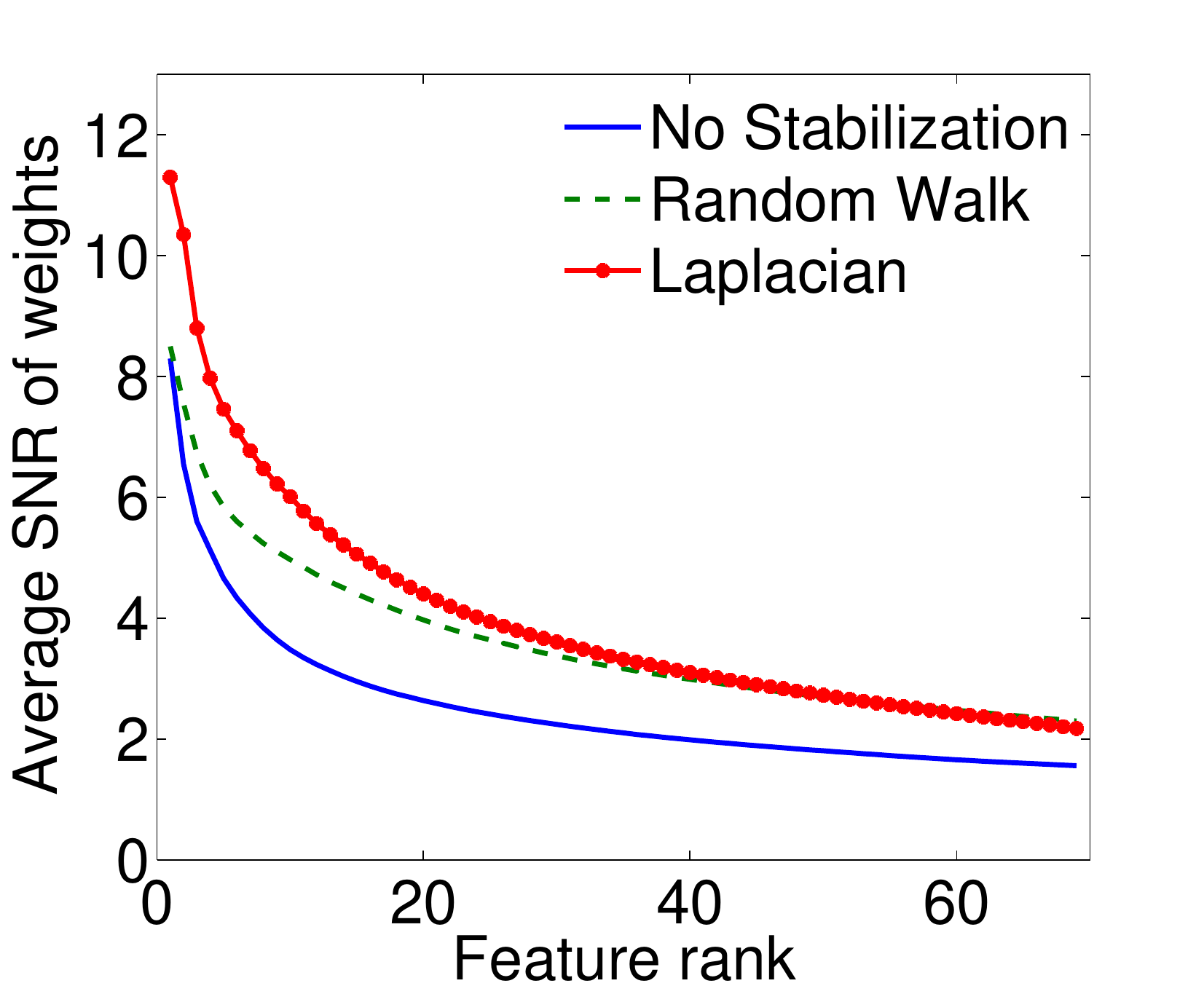}\tabularnewline
(a) Cumulative, $C_{2}+C_{3}$ & (b) Stagewise (Shared), $C_{2}+C_{3}$\tabularnewline
\includegraphics[width=0.4\textwidth]{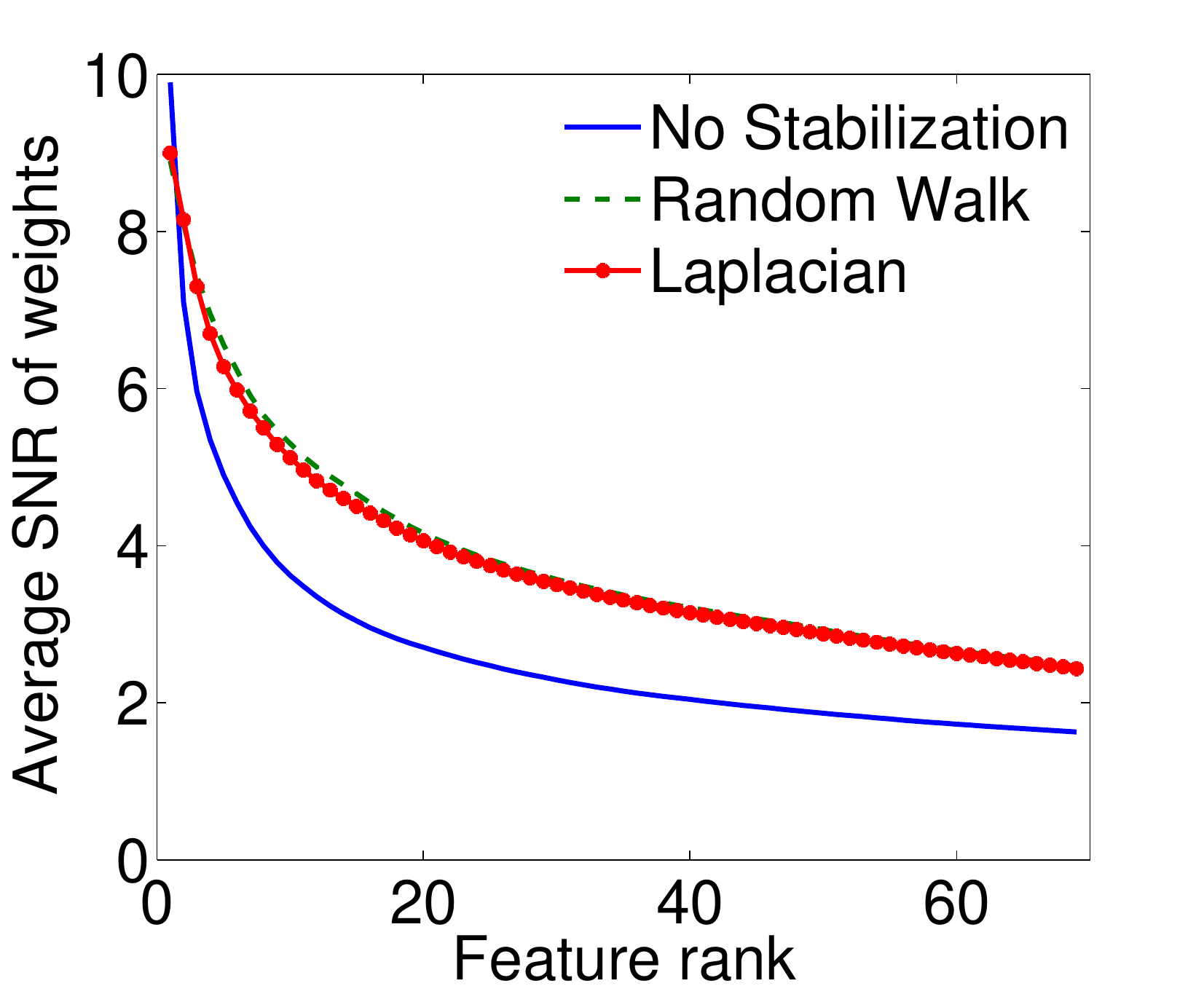} & \includegraphics[width=0.4\textwidth]{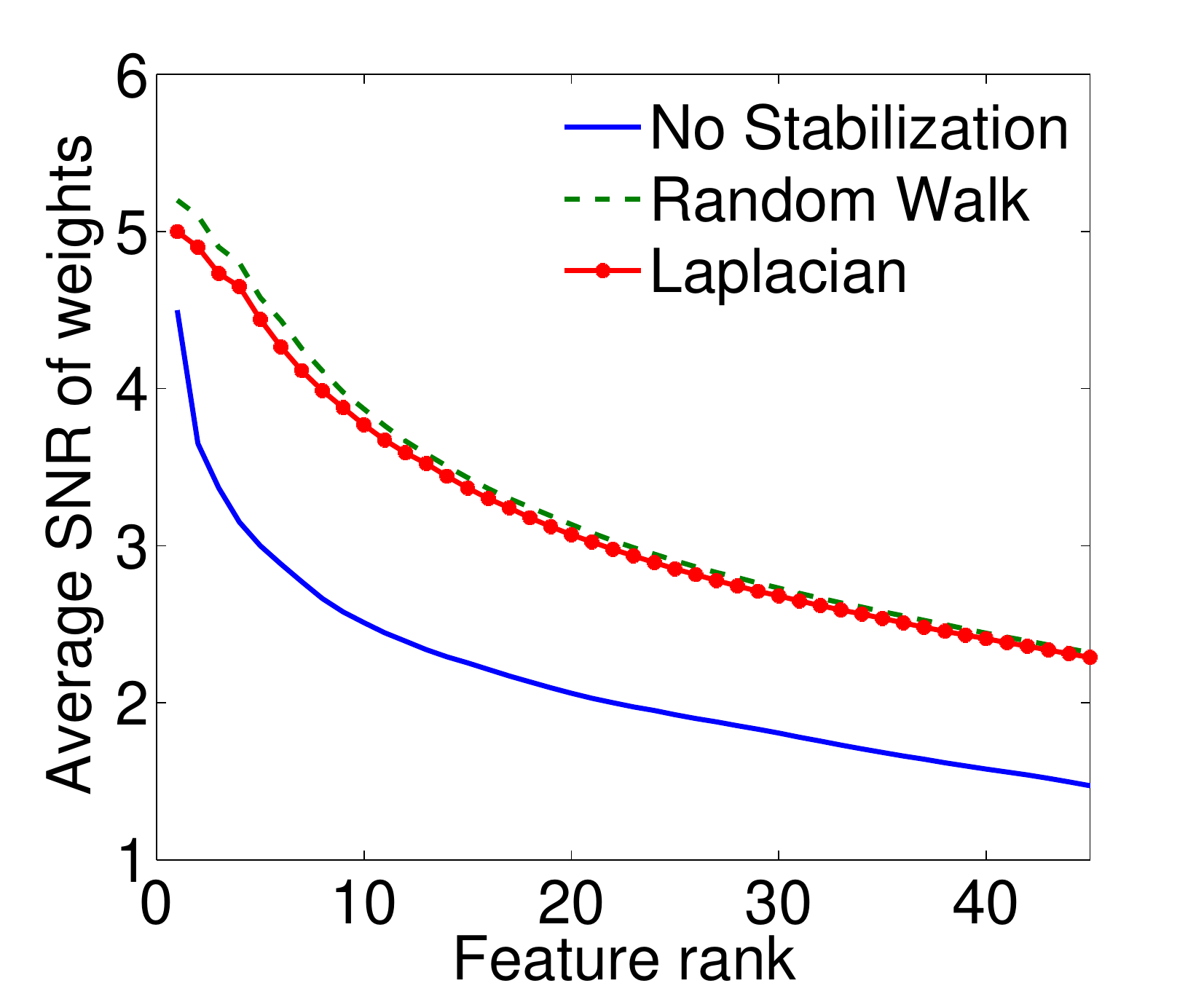}\tabularnewline
(c) Stagewise (Separate), $C_{2}$ & (d) Stagewise (Separate), $C_{3}$\tabularnewline
\end{tabular}
\par\end{centering}

\caption{Average signal-to-noise ratio ($SNR@T$-- see Eq.~(\ref{eq:SNR-index})),
evaluated at different rank list sizes (the larger average SNR, the
more stable model). ``No Stabilization'' means the standard lasso
framework with $\beta=0$. For others, $\alpha=3\times10^{-4},\beta=3\times10^{-3}$.
\label{fig:Average-SNR}}
\end{figure}

We now examine the models stability against data sampling and evaluate
the stabilizing property of the proposed method (Sec.~\ref{sub:Stablizing-Sparse-Models}).
For each fold, we generated $30$ samples, each of which was drawn
randomly from $50$\% of training data. Each example resulted in a
model, and the feature weights were recorded and finally the results
of all $10$ folds -- $300$ models -- were combined. Figs.~\ref{fig:Average-selection-probability}(a--d)
show the $ASP@T$ indices (Eq.~(\ref{eq:ASP-index})) as functions
of the rank list size $T$, for all ordinal classifiers. The instability
is clearly an issue -- the average selected probability drops as more
features are included. Using both the Laplacian and random walk regularization
methods (Eqs.~(\ref{eq:Laplacian},\ref{eq:random-walk})), the improvement
in stability is evidenced in all settings. The instability and stabilizing
effect were similarly obtained with the $SNR@T$ indices (Figs.~\ref{fig:Average-SNR}(a--d)).

\subsection{Discovered Features}

\begin{table}
\begin{centering}
\begin{tabular}{lccc}
\hline 
\emph{Feature} & ($\sigma_{k};s_{k}$) & \emph{Importance} & \emph{SNR}\tabularnewline
\hline 
Number of EDs & ($3;0$) & 59.6 & 8.5\tabularnewline
Number of EDs & ($3;3$) & 32.5 & 6.6\tabularnewline
Moderate-lethality attempts ($C_{2}$) & ($12;12$) & 10.9  & 5.1 \tabularnewline
Moderate-lethality attempts ($C_{2}$) & ($6;6$) & 9.1  & 4.1 \tabularnewline
High-lethality attempts ($C_{3}$) & ($3;0$) & 14.6 & 4.4 \tabularnewline
Moderate-lethality attempts ($C_{2}$) & ($24;24$) & 12.1  & 4.4 \tabularnewline
Number of EDs & ($6;6$) & 22.6 & 4.3\tabularnewline
Number of postcode changes \& Male & ($3;0$) & 26.0 & 4.0 \tabularnewline
High-lethality attempts ($C_{3}$) & ($3;3$) & 10.3 & 4.0 \tabularnewline
ICD code: \emph{F19} (Mental disorders due to drug abuse)  & ($6;6$) & 9.8 & 3.8\tabularnewline
ICD code: \emph{Z91} (History of risk-factors, unclassified) & ($24;24$) & 9.0 & 3.6\tabularnewline
High-lethality attempts ($C_{3}$) & ($6;6$) & 8.0 & 3.3 \tabularnewline
ICD code: \emph{T50} (Poisoning)  & ($3;0$) & 14.6  & 3.2 \tabularnewline
ICD code: \emph{Z29} (Need for other prophylactic measures) & ($3;0$) & 24.0 & 3.2 \tabularnewline
Number of postcode changes \& Male & ($3;3$) & 10.8 & 3.0 \tabularnewline
Comorbidity: Alcohol abuse & ($6;6$) & 5.2 & 2.9\tabularnewline
Number of EDs & ($12;12$) & 12.6 & 2.8\tabularnewline
ICD code: \emph{S06} (Intracranial injury) & ($3;0$) & 2.9 & 2.7 \tabularnewline
ICD code: \emph{U73} (Other activity)  & ($3;0$) & 6.8 & 2.5 \tabularnewline
ICD code: \emph{T43} (Poisoning by psychotropic drugs) & ($3;0$) & 7.4 & 2.5 \tabularnewline
\hline 
\end{tabular}
\par\end{centering}

\caption{Top $20$ predictive and stable features associated with risky outcomes
in the next $3$ months, ranked by \emph{signal-to-noise ratio}s,
as produced by the stagewise classifier with shared parameters (Sec.~\ref{sub:Stagewise-Models}),
under Laplacian regularization (Eq.~\ref{eq:Laplacian}). The uniform
kernel width $\sigma_{k}$ and the delay $s_{k}$ are measured in
months; \emph{ED} = Emergency Attendance, \emph{MHDG} = Mental Health
Diagnosis Group. \label{tab:Top-ranked-features}}
\end{table}

Cumulative classifiers and stagewise classifiers with shared parameters
do not distinguish the parameters between classes and thus we have
a single list of features at the end of the training phase. Tab.~\ref{tab:Top-ranked-features}
presents top $20$ features ordered by their SNRs, as produced by
the stagewise classifier with shared parameters (Sec.~\ref{sub:Stagewise-Models}).
Predictive features include: Recent emergency visits, recent high-risk
attempts ($C_{3}$), moderate-risk attempts ($C_{2}$ \& self-poisoning)
within 24 months, recent history of mental problems and of drug abuse,
socioeconomic problems (frequent home moving). 

\begin{table}
\begin{centering}
\begin{tabular}{llccc}
\hline 
\multicolumn{2}{l}{\emph{Feature}} & ($\sigma_{k};s_{k}$) & \emph{Importance} & \emph{SNR}\tabularnewline
\hline 
\multicolumn{2}{l}{\textbf{Moderate-risk class} $(C_{2})$} &  &  & \tabularnewline
 & Number of EDs & ($3;0$) & 59.1  & 9.0\tabularnewline
 & Number of EDs & ($3;3$) & 32.9  & 7.3\tabularnewline
 & Moderate-lethality attempts ($C_{2}$) & ($6;6$) & 13.6  & 5.6 \tabularnewline
 & Moderate-lethality attempts ($C_{2}$) & ($12;12$) & 13.7 & 4.9\tabularnewline
 & Number of EDs & ($6;6$) & 18.3 & 4.6 \tabularnewline
 & Moderate-lethality attempts ($C_{2}$) & ($24;24$) & 15.6 & 4.3\tabularnewline
 & ICD code: \emph{Z91} (History of risk-factors, unclassified) & ($24;24$) & 8.5 & 4.1\tabularnewline
 & Comorbidity: Alcohol abuse & ($6;6$) & 7.0 & 4.1\tabularnewline
 & Moderate-lethality attempts ($C_{2}$) & ($3;3$) & 9.4  & 3.6 \tabularnewline
 & Comorbidity: Alcohol abuse & ($3;3$) & 8.1 & 3.6\tabularnewline
 & Number of postcode changes \& Male & ($3;0$) & 20.0 & 3.4 \tabularnewline
 & ICD code: \emph{Z29} (Need for other prophylactic measures) & ($3;0$) & 28.2 & 3.3 \tabularnewline
 & Moderate-lethality attempts ($C_{2}$) & ($3;0$) & 9.7 & 3.3 \tabularnewline
 & Comorbidity: Alcohol abuse & ($3;0$) & 9.0 & 3.2\tabularnewline
 & ICD code: \emph{F19} (Mental disorders due to drug abuse)  & ($6;6$) & 7.8 & 3.1\tabularnewline
\multicolumn{2}{l}{\textbf{High-risk class} $(C_{3})$} &  &  & \tabularnewline
 & ICD code: \emph{T43} (Poisoning by psychotropic drugs) & ($3;0$) & 24.0 & 5.0 \tabularnewline
 & High-lethality attempts ($C_{3}$) & ($3;0$) & 30.6 & 4.8 \tabularnewline
 & ICD code: \emph{T43} (Poisoning by psychotropic drugs) & ($3;3$) & 15.2 & 4.1 \tabularnewline
 & High-lethality attempts ($C_{3}$) & ($3;3$) & 24.0 & 4.1 \tabularnewline
 & ICD code: \emph{T42} (Poisoning by antiepileptic,  & ($3;0$) & 15.3 & 3.6\tabularnewline
 & sedative-hypnotic and antiparkinsonism drugs) &  &  & \tabularnewline
 & ICD code: \emph{U73} (Other activity)  & ($3;3$) & 10.6 & 3.4\tabularnewline
 & ICD code: \emph{T42} (Poisoning by antiepileptic, & ($3;3$) & 11.0 & 3.2\tabularnewline
 &  sedative-hypnotic and antiparkinsonism drugs) &  &  & \tabularnewline
 & ICD code: \emph{T50} (Poisoning)  & ($3;0$) & 17.3  & 3.1 \tabularnewline
 & ICD code: \emph{X61} (Intentional self-poisoning) & ($3;0$) & 10.0 & 3.0\tabularnewline
 & Occupation: student \& Female & \emph{NA} & 69.5 & 2.8\tabularnewline
 & High-lethality attempts ($C_{3}$) & ($6;6$) & 13.3 & 2.7 \tabularnewline
 & ICD code: \emph{U73} (Other activity)  & ($3;0$) & 11.4 & 2.7\tabularnewline
 & ICD code: \emph{X61} (Intentional self-poisoning) & ($3;3$) & 6.3 & 2.7\tabularnewline
\hline 
\end{tabular}
\par\end{centering}

\caption{Top $15$ predictive and stable features associated with risk classes
in the next $3$ months, ranked by \emph{signal-to-noise ratio}s,
as produced by the stagewise classifier \emph{without} parameter sharing
(Sec.~\ref{sub:Stagewise-Models}). The uniform kernel width $\sigma_{k}$
and the delay $s_{k}$ are measured in months; \emph{MHDG} = Mental
Health Diagnosis Group.\label{tab:Top-ranked-features-C2-C3}}
\end{table}

Stagewise classifiers with class-specific parameters can offer re-ranking
of features for $C_{2}$ and $C_{3}$ separately. Tabs.~\ref{tab:Top-ranked-features-C2-C3}
list top-ranked class-specific features for $C_{2}$ and $C_{3}$,
respectively, under the stagewise classifiers. A noticeable aspect
is the strong association between prior $C_{3}$ attempts with future
$C_{3}$ outcomes.

\section{Discussion \label{sec:Discussion}}

Compared against existing work in medical risk models, our machine
learning method is hypothesis-free, i.e., without collection bias
nor prior assumptions about specific risk factors. As the model is
derived from routinely collected administrative hospital data, it
can be readily embedded into existing EMR systems. Second, as all
available information is utilized, there is less chance that importance
risk factors will be overlooked. In fact, the features we have just
discovered (Tables~\ref{tab:Top-ranked-features} and \ref{tab:Top-ranked-features-C2-C3})
resemble ones well-documented in the clinical literature \cite{brown2000risk}.
For example, male, socio-economic issues, psychiatric factors, previous
attempts are known to be positively correlated with subsequent suicide
attempts \cite{gonda2012suicidal,haw2011living,martin2012clinical}.
Factors distantly related to psychological distress such as prior
hospitalization, ED visits or physical illnesses were also previously
reported \cite{da2011emergency,luoma2002contact,qin2013hospitalization}.
Our results, however, differentiate from the previous work since they
are more precise in timing, and do not rely on hand-crafted prior
hypotheses. 

The result is significant as the discovery is essentially free and
automated, as compared to expensive and time-consuming medical studies.
However, as EMR data may contain noise and depend on system implementation,
discovered risk factors may not totally universal, and our system
thus should be used as a fast screening tool for further in-depth
clinical investigation. 

Unlike existing machine learning work applied to healthcare, our goal
was to achieve not only high performance but also interpretability and reproducibility.
The prediction is transparent and for each patient, it explains specific risk factors
involved in the risk estimate, and how stable the risk factors are
(Tables~\ref{tab:Top-ranked-features}~and~\ref{tab:Top-ranked-features-C2-C3}).
Although feature stability has gained significant attention recently \cite{austin2004automated,kalousis2007stability},
this has not been studied in the context of clinical prediction models.
Further, we contribute to the literature two new stability indices,
the \emph{ASP@T} and \emph{SNR@T, }where the \emph{SNR@T} measures
not only the feature stability but also its statistical significance
\emph{(}the Wald statistic). This statistic is largely ignored in
data mining practice. 

Our work demonstrated that model stability for high-dimensional problems
could be significantly enhanced by exploiting known relations between
features. This validates our intuition that prior knowledge would
help as it is independent of data sampling procedures. Consistent
with prior studies, our results confirm that such prior relations,
as realized in feature network regularization, improve the generalization
when no other regularization schemes are in place \cite{fei2010regularization}\cite{sandler2008regularized}.
However, interestingly, when combined with lasso, their effect on
predictive performance is insignificant, as shown in Fig.~\ref{fig:Performance-versus-hyperparameters}.
It is surprising because model stability could potentially lead to
better prediction stability, and which is a sufficient condition for
generalization \cite{bousquet2002stability,poggio2004general}. This
suggests that the two stability concepts may not be strongly correlated,
as it is known that random forests, for example, can generate very
different tree ensembles (model instability) but the end results can
be quite stable (prediction stability).

This paper grew from an effort to predict suicide, following difficulty
in practice at Barwon Health, Australia. However, this goal was quickly
deemed impossible, partly because of the long-standing conjecture
that suicide is clinically unpredictable \cite{large2010suicide,large2012suicide}.
From the machine learning perspective, suicide is a rare event, and thus
a robust estimation would require detailed clinical data from millions
of mental health patients, which is an impractical task. Instead,
the mental health literature has concentrated on predicting suicide
attempts without stratifying lethality. However, it is possible that
the mental processes in low-lethality attempts differ significantly
from those in high-lethality attempts, as suggested in Table~\ref{tab:Top-ranked-features-C2-C3}.
In practice, clinicians would want to target the latter group because
of the high chance of subsequent death. Thus our paper contributes
to the literature by separating the lethality classes in an ordinal
regression framework. The lesson is that when facing rare events,
instead of predicting the events itself, we should target regions
where the events are likely to occur. Another finding is that, in
medical domains, machine learning systems are most useful when data
is large, comprehensive and complex, the risk factors are abundant,
time-sensitive but weakly predictive. This is because clinicians,
in their busy practice, may not able to consider a large number of
relevant factors in the distant history.

This study has several limitations. First the framework has only been
validated on data from a single hospital and has not been independently
tested by external investigators. However, the framework has been
tested on a variety of medical problems and cohorts (results are reported
elsewhere), and the results so far have been encouraging: The predictive
performance either matches or exceeds the state-of-the-arts in the
clinical literature, and the discovered features resembles most important
reported risk factors from multiple prior studies. Second, our labels
were not perfect: (i) labels were collected at Barwon Health alone
and we did not track transfers or readmissions to other institutions;
(ii) labels were based on ICD-10 diagnosis codes and these may not
be perfectly accurate due to the coding practices. This suggests that
our performance estimates are conservative.

\section{Conclusion \label{sec:Conclusion}}

We have proposed a stabilized sparse ordinal regression framework
for future risk stratification. The objectives are deriving and validating
predictive algorithms from the rich source of electronic health records,
and at the same time, offering clear explanation on how prediction
is made. Central to the work is discovery of stable subset of factors
that are predictive of future risk. The framework has several novel
elements: (i) two model stability indices; (ii) a stability-based
feature ranking criterion; and (iii) feature network regularization
where similar features are encouraged to have similar weights, under
the lasso-based sparsity framework. 

The framework introduced in this paper is generalizable as the information
extracted from the data warehousing is standardized. The EMR-based
models make no use of human resources, except for the risk definition
done only once. Our framework has been validated on a challenging
problem of predicting suicide risk against clinicians. We demonstrated
in this paper that the proposed system could (a) discover risk factors
that are consistent with mental health knowledge; (b) significantly
outperform clinicians using just readily collected data in hospitals;
and (c) exploit feature relations improved model stability significantly. 

Work in progress is testing the framework on a series of other predictive
problems: Risk of hospitalization/mortality in diabetes, stroke, COPD,
mental health, heart failure, heart attack and pneumonia, and cancers.
The framework has been adopted by the hospitals and deployment is
underway. This poses an interesting research question: How can we
deal with the situation where the physicians modify their treatment
strategy based on the machine prediction, and thus alter the outcome,
leading to the poorer match between the actual outcome and the predicted
one?

\subsection*{Acknowledgments}

We thank Ross Arblaster and Ann Larkins for helping data collections,
Paul Cohen for providing management support for the project, Richard
Harvey for risk stratification, Michael Berk and Richard Kennedy for
valuable opinions, and the reviewers for helpful comments.


\end{document}